\documentclass{article}
\usepackage[a4paper,margin=1in]{geometry}
\usepackage{graphicx} 
\graphicspath{{Figures/}}
\usepackage{placeins}
\usepackage{authblk}
\usepackage{comment}
\usepackage{booktabs}
\usepackage{amsmath}
\usepackage{mathtools}
\usepackage{amssymb}
\usepackage{tabularx}
\newcolumntype{Y}{>{\centering\arraybackslash}X}
\usepackage{xcolor}
\usepackage{array}
\usepackage[numbers]{natbib}
\usepackage{appendix}

\newcommand{\y}{\pmb{\mathrm{y}}}
\newcommand{\Y}{\pmb{\mathrm{Y}}}
\newcommand{\tramp}{t_{\text{ramp}}}
\newcommand{\kramp}{K_{\text{ramp}}}
\newcommand{\vor}{\pmb{\omega}}
\newcommand{\pres}{\pmb{C}_p}
\newcommand{\lift}{C_L}
\newcommand{\drag}{C_D}
\newcommand{\ang}{\alpha}
\newcommand{\angv}{\dot{\alpha}}
\newcommand{\lat}{\pmb{\xi}}
\newcommand{\Lat}{\pmb{\Xi}}
\newcommand{\weights}{\pmb{W}}
\newcommand{\h}[1]{\hat{#1}}
\newcommand{\force}{\pmb{F}}
\newcommand{\encoder}{\mathcal{F}_e}
\newcommand{\decoder}{\mathcal{F}_d}
\newcommand{\obs}{\pmb{h}_{\weights}}
\newcommand{\frw}{\pmb{f}_{\weights}}

\newcommand{\obsnoise}{\pmb{\eta}}

\newcommand{\frwnoise}{\pmb{w}}
\newcommand{\gain}{\pmb{K}}
\newcommand{\stateCov}{\pmb{\Sigma}_{\lat}}
\newcommand{\obsCov}{\pmb{\Sigma}_{\text{y}}}
\newcommand{\obsstd}{\sigma_{\text{y}}}
\newcommand{\stateObsCov}{\pmb{\Sigma}_{\tilde{\Lat} \tilde{\Y}}}
\newcommand{\obsObsCov}{\pmb{\Sigma}_{\tilde{\Y} \tilde{\Y}}}

\newcommand{\state}{\pmb{\mathrm{x}}}
\newcommand{\Obs}{\pmb{H}}
\newcommand{\sigmax}{\pmb{\Sigma}_{\pmb{\mathrm{x}}}}

\title{A learning-based joint flow and kinematic state estimation for bodies in highly disturbed flows}
\author[1,*]{Hanieh Mousavi}
\author[1]{Anya Jones}
\author[1]{Jeff Eldredge}
\affil{Mechanical and Aerospace Engineering, University of California, Los Angeles, Los Angeles, CA 90095-1597, USA}
\affil[*]{Corresponding author, email: hnmousavi@ucla.edu}
\date{}

\begin{document}

\maketitle

\begin{abstract}
    Accurate estimation of unsteady aerodynamic flows from sparse surface measurements remains a fundamental challenge, particularly in the presence of strong disturbances, unknown body kinematics, and incomplete observations. This study presents a data-driven sequential estimation framework for the joint reconstruction of unsteady flow fields, aerodynamic loads, and airfoil kinematics from sparse surface pressure measurements. The proposed approach combines a kinematics-aware nonlinear flow autoencoder for efficient data compression with an online filtering strategy that assimilates streaming measurements. Within the resulting reduced-order representation, the forecast and observation operators---two key components required by the filtering framework---are learned directly from data. Each forecast–assimilation cycle incurs only a few milliseconds, highlighting the potential of the proposed reduced-order framework for real-time aerodynamic state estimation.
    The framework is trained and evaluated on computations of two-dimensional incompressible flow over an airfoil subjected to random vortical gusts while undergoing arbitrary pitch-up motions. The results demonstrate accurate reconstruction of transient flow structures, aerodynamic loads, and kinematic states, from a variety of classes of limited measurement: a small number of surface pressure sensors, vertical lines of vorticity sensors, and synthetic velocimetry data with representative shadow regions. The transient informativeness of individual pressure sensors during the disturbance–airfoil interaction is quantified through their time-varying contributions to the dominant observation modes; it is shown that, while lift is reconstructed from the leading modes, drag, pitch angle, and angular velocity require higher observation modes. Incorporating limited off-body measurements alongside surface pressure data is shown to further improve observability and leads to reduced estimation error and uncertainty. 
    
\end{abstract}

\section{Introduction}\label{sec:introduction}
Accurate characterization of unsteady flow fields around moving bodies remains a central challenge in the fields of unsteady aerodynamics and fluid–structure interaction (FSI), with direct implications for flow control, load mitigation, and performance prediction. Even when body motion is prescribed---such as wing pitching, plunging, or flapping---the resulting flow evolution and aerodynamic loads depend not only on the instantaneous kinematic state but also on its temporal history, often through strongly nonlinear and nonlocal mechanisms \citep{chiereghin2019unsteady, kurtulus2019unsteady}. These challenges are particularly acute in practical settings involving gust load alleviation \citep{fukami2023grasping, mousavi2025low, fukami2025extreme}, maneuvering flight, bio-inspired propulsion \citep{de2024bio}, and sensing-based control \citep{renn2022machine}, where real-time access to reliable estimates of both flow and motion states and aerodynamic forces is essential for decision making \citep{mousavi2025low, mousavi2025sequential, liu2025attention}.
In such applications, however, full-field flow measurements are rarely available. Experimental and operational constraints, moving boundaries, optical occlusions, and the need for real-time sensing often limit observations to sparse, indirect measurements, such as surface-mounted pressure sensors. As a result, estimation frameworks must infer high-dimensional, transient flow states to enable load reconstruction and downstream flow control, from limited and noisy data, often in the presence of strong disturbances.

Nature demonstrates that this task is not fundamentally intractable. Biological swimmers and flyers extract actionable flow information using distributed pressure and velocity cues along their bodies, enabling perception and control using only local sensing \citep{bleckmann2009lateral, mogdans2012coping}. Inspired by these mechanisms, a growing body of work has explored data-driven inference of flow fields and coupled system states from sparse observations. For example, Tang et al.~\cite{tang2025neural} inferred coupled fluid–structure states from limited off-body measurements, while Zhu et al.~\cite{zhu2025physics} reconstructed flow fields and hidden boundaries using physics-informed learning. Focusing on surface-based sensing, Rodwell et al.~\cite{rodwell2024feel} reconstructed unsteady flow fields from local pressure measurements through learned modal representations using windowed data segments, and Liu et al.~\cite{liu2025attention} leveraged histories of sparse pressure data for reinforcement-learning-based control of a pitching airfoil.

Despite these advances on estimating the flow field around moving/deforming bodies, three critical limitations remain unresolved.
First, most existing frameworks assume either a fixed body or a body with known motion histories, thereby focusing exclusively on estimating the flow field. Second, many approaches rely on offline reconstruction, impeding their applicability to online estimation and control. Third, the majority of methods implicitly assume complete or uniformly observed flow fields---an assumption that is routinely violated in experimental and operational environments. 

A particularly severe manifestation of this last limitation arises in experiments with spatially incomplete flow measurements. In planar particle image velocimetry (PIV), measurements are often unavailable in localized regions due to laser-sheet blockage, line-of-sight restrictions, or physical interference from the body or mounting hardware. These shadowed regions are ubiquitous in wind-tunnel and water-channel experiments and are typically excluded from reported results, as seen in studies of pitching airfoils \citep{rezapour2026dynamic}, fixed-angle airfoils under transverse gusts \citep{gementzopoulos2025flow}, and rotating or pitching lifting surfaces \citep{medina2016leading}. \citet{gementzopoulos2025flow} have shown that even flow regions lacking strong vortical activity---such as attached or weakly separated flow on the pressure side of an airfoil---can exert a measurable influence on surface pressure distributions and integrated aerodynamic loads. Their incomplete observation therefore poses a fundamental challenge for flow-state estimation, reduced-order modeling, and data-assimilation methods.
Several studies have investigated the reconstruction of experimentally missing flow regions using variants of gappy proper orthogonal decomposition (POD), originally introduced by \citet{everson1995karhunen}. These include spectral extensions that leverage temporal coherence, such as gappy spectral POD \citep{nekkanti2023gappy}, as well as related modal approaches developed specifically for PIV measurements with occluded regions \citep{venturi2004gappy}. More recently, machine-learning-based methods---ranging from autoencoders to other neural-network architectures---have been explored for reconstructing spatially incomplete flow fields using inpainting approaches \citep{luo2023reconstruction, aksoy2023reconstruction, luo2024deep}. Although effective for post-processing and offline reconstruction, these methods are generally not formulated for online, sequential estimation, nor do they naturally account for nonlinear, transient disturbance dynamics or uncertainty quantification.

Taken together, these limitations point to a fundamental gap in the current literature: there is presently no framework capable of \emph{online and in real time} estimation of unsteady flow fields, aerodynamic loads, and time-varying body kinematics in a strongly disturbed environment with nonlinear interactions, using only sparse measurements.
Addressing this gap is essential for enabling closed-loop sensing, control, and decision-making in gusty environments.

The objective of the present study is thus to introduce a fast, data-driven estimation framework that enables simultaneous reconstruction of unsteady flow fields, aerodynamic loads, and body kinematics for bodies undergoing arbitrary motions in strongly disturbed flows, using sparse, noisy measurements. The proposed framework is designed to operate in high-dimensional settings and yet remain suitable for online deployment and employ realistic sensor suites.
While our previous work \citep{mousavi2025sequential} addressed disturbed-flow estimation for a fixed airfoil, the present study considers a substantially more challenging inverse problem in which the body motion itself is unknown and must be inferred together with the surrounding flow states. In this setting, surface pressure measurements contain coupled signatures of incoming disturbances, body motion, and unsteady aerodynamic phenomena, making the estimation problem fundamentally different from that of a stationary body.
To address this challenge, we introduce a kinematics-aware nonlinear reduced-space representation---observable from sparse measurements---that jointly encodes the aerodynamic and kinematic states of the system. To the authors' knowledge, this is the first reduced-order sequential estimation framework capable of jointly reconstructing aerodynamic states and time-varying body kinematics in strongly disturbed flows. The present study provides an intermediate yet essential foundation for addressing fully coupled FSI problems.
The present work also investigates the time-varying observability of states of interest through sparse surface pressure measurements. By examining the contributions of individual pressure modes to the estimation of lift, drag, pitch angle, and angular velocity, the study provides new insight into the sensing requirements of moving bodies in unsteady environments.
Finally, we investigate two complementary approaches of multi-modal sensing for improving estimation in weakly observable regions. The first incorporates limited off-body measurements in the form of vorticity measurements along a small number of vertical slices to assess how strategically placed sensors improve estimation accuracy. The second considers experimentally relevant situations involving spatially incomplete observations and demonstrates the ability of the estimator to reconstruct dynamically important flow structures within shadowed regions of the flow domain.

The remainder of the paper is organized as follows. Section~\ref{sec:problem} formulates the problem and presents the mathematical framework for a pitching-up airfoil subjected to transient strong disturbances. Section~\ref{sec:results} reports results under transient disturbances, demonstrating accurate and efficient online state estimation. Section~\ref{sec:conclusion} summarizes the main findings and outlines limitations and directions for future work.

\section{Problem description and methodology}\label{sec:problem}
\begin{figure*}
\centering
\includegraphics[width=1.0\textwidth]{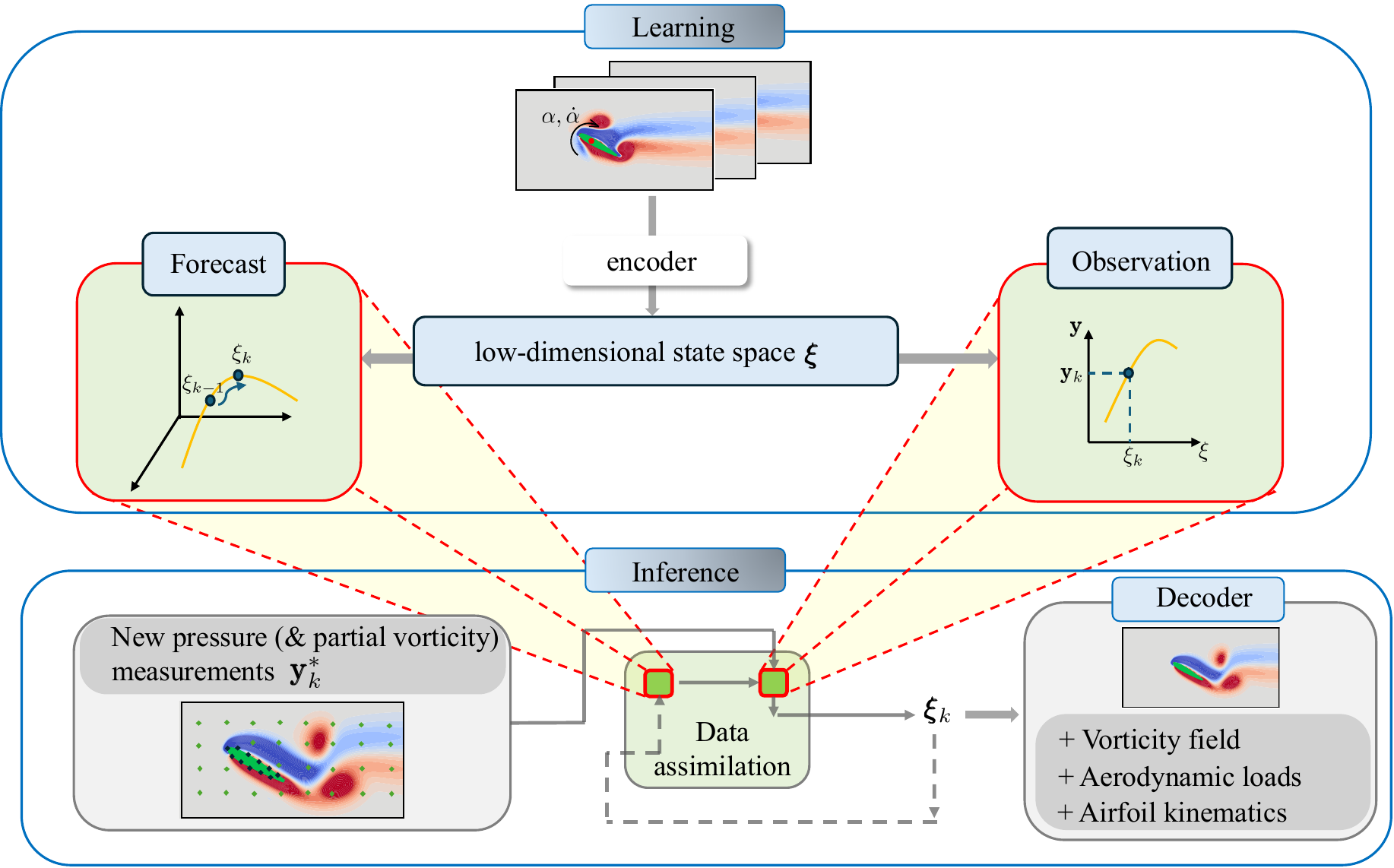}
\caption{\label{fig:diagram} Block diagram summarizing the framework developed and applied in this study. The framework is extended from \cite{mousavi2025sequential} to jointly estimate aerodynamic flow states and body kinematics for an airfoil undergoing unknown ramp pitch-up maneuvers. The learning stage is performed offline, whereas state inference is carried out online.}
\end{figure*}

In this section, we formulate the problem and outline the methodology adopted to address it. The overarching objective of the present study is to enable fast sequential estimation of high-dimensional flow and motion kinematic states from sparse and noisy measurements. Building on our recent work \citep{mousavi2025sequential}, which focused on a fixed airfoil configuration, we extend the framework here to accommodate a pitching airfoil. To make filtering computationally tractable in this setting, we learn---offline---a low-dimensional latent state space and the corresponding reduced-order models required by a data assimilation framework, and then perform the online filtering step entirely in the learned low-dimensional space \citep{kaveh2026data}. 
A schematic overview of the proposed pipeline is shown in Fig.~\ref{fig:diagram}. We first train an autoencoder to discover a compact latent manifold, denoted by $\lat$, that captures the dominant variability of the high-dimensional flow fields and the body kinematics. The forecast (dynamical) and observation operators are learned in this latent space. Together, these two components provide the reduced-order counterparts of the prediction and measurement models required in the filtering framework.

During online inference, the latent-state probability is propagated forward with the learned forecast model to obtain a prior (forecast) estimate. As measurements $\y_k^*$ arrive at time step $k$, the filtering update is performed in the latent space to infer the posterior reduced state $\lat_k$ consistent with the measured data and their uncertainty, and this loop is carried out recursively. At any desired point, the posterior samples $\lat_k$ can be passed through the decoder to reconstruct the corresponding full flow field and derived quantities of interest, thereby producing computationally-efficient state estimates together with uncertainty bounds in the original high-dimensional space. Details of the learned operators and the sequential filtering procedure are deferred to Section~\ref{sec:filtering}.

\subsection{Problem statement}
\begin{figure*}
\centering
\includegraphics[width=0.8\textwidth]{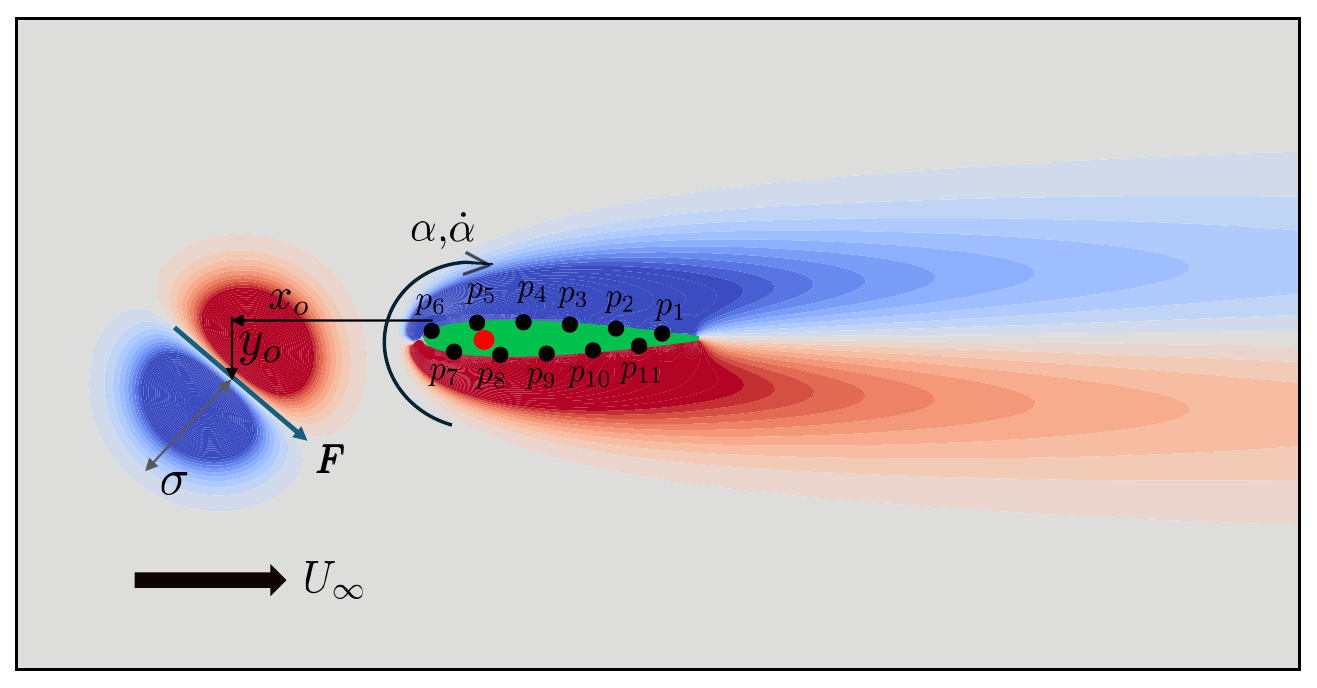}
\caption{\label{fig:configuration} Configuration of the problem, illustrating the relative position of the gust centre with respect to the leading edge of the airfoil, the size of the Gaussian disturbance and the indices of sensors mounted on the airfoil. The airfoil is pitching up about the quarter chord (red circle) with a smooth ramp motion.}
\end{figure*}

The framework must be trained on data obtained from the class of flows of interest. For the purposes of this paper, we simulate two-dimensional incompressible flow of a fluid with density $\rho$ and kinematic viscosity $\nu$ with uniform velocity $U_\infty$ over a NACA 0012 airfoil with chord length $c$ and instantaneous angle of attack $\ang$. Throughout, we assume an undisturbed chord-based Reynolds number of $Re=U_\infty c / \nu = 100$. A random disturbance is introduced upstream of the airfoil via a body force $\force$, Gaussian in space and time, applied to the fluid as
\begin{equation}
    \force(x,y,t) = \rho U_\infty c^2 \frac{(D_x,D_y)}{\pi^{3/2} \sigma_x \sigma_y \sigma_t} \exp \left[ - \frac{(x-x_o)^2}{\sigma_x^2} - \frac{(y-y_o)^2}{\sigma_y^2} - \frac{(t-t_o)^2}{\sigma_t^2} \right],
\end{equation}
where $(D_x,D_y)$ are the dimensionless forcing amplitudes in the streamwise and transverse directions, $\sigma_x$ and $\sigma_y$ are the spatial spreads of the forcing field, and $\sigma_t$ is its temporal width. The gust center is located at $(x_o,y_o)$ relative to the leading edge (LE) of the airfoil (placed at the origin) and introduced at time $t_o$. The schematic of the problem configuration is illustrated in figure~\ref{fig:configuration}. In all simulations, we fix $D_x=2.0$, $x_o=-0.8c$, and $\sigma_t=0.1c/U_\infty$, while other gust parameters are randomly sampled from the ranges $D_y \in [-2.0,2.0]$, $\sigma_x=\sigma_y=\sigma \in [0.05c,0.2c]$, $y_o \in [-0.25c,0.25c]$, and $t_o \in [0.3c/U_\infty,2.0c/U_\infty]$.

The airfoil undergoes a prescribed smooth ramp pitch-up and hold motion about the quarter-chord, defined using the smooth ramp function \citep{eldredge2009computational}
\begin{equation}
    \ang(t) = \ang_o + \frac{\dot{\alpha}_o}{2a_s} \ \ln \left( \frac{\cosh \left(a_s(t-\tramp) \right)}{\cosh \left(a_s(t-\tramp - \Delta \tramp) \right)} \right) + \frac{\Delta \ang}{2},
\end{equation}
where $\Delta \tramp = \Delta \ang / \dot{\alpha}_o$ denotes the duration of the ramp motion, $\Delta \ang$ is the total pitch increment, $\dot{\alpha}_o$ is the nominal pitch rate, $\tramp$ is the onset time of the ramp, $\ang_o$ is the initial angle of attack, and $a_s$ is a smoothness parameter that controls the sharpness of the transition at the beginning and end of the motion. In the present study, the value of $a_s=11$ is used. For all moving-body cases considered here, the airfoil is rotated about the quarter chord from an initial angle of $\ang_o=0$ to a final angle of $\ang_{\max} =  \Delta \ang = \pi/6 = 30^\circ$, with the ramp initiated at $\tramp=1$. The nominal pitch rate $\dot{\alpha}_o$ is described, as is typical \cite{eldredge2009computational}, by the dimensionless parameter $\kramp = \dot{\alpha}_o c/(2 U_\infty)$. This dimensionless parameter is randomly selected from a range of values $\kramp \in [0.125, 1.25]$, yielding a range of ramp durations. Such rate-limited pitching motions are representative of practical control-induced variations in angle of attack, where actuator limitations and stability considerations preclude impulsive maneuvers. These kinematics are widely used in studies of unsteady aerodynamics and dynamic stall \citep{eldredge2009computational, eldredge2019leading, cavanagh2024effect}. 

To explore the high-dimensional space spanned by the gust and ramp parameters, we employ Latin Hypercube Sampling (LHS) to construct a diverse ensemble of realizations. This stratified sampling strategy provides near-uniform coverage of the prescribed parameter ranges while substantially reducing the number of simulations needed for training and evaluation. Here we generate 100 independent cases of random combinations of gust and ramp.

The incompressible Navier–Stokes equations are solved in vorticity–streamfunction form using the Lattice Green’s Function / Immersed Layers method developed by \citet{eldredge2022method}. All simulations are performed on a Cartesian computational domain extending over $(-2c, 4c) \times (-2c, 2c)$ in the $x$ and $y$ directions, respectively, with a uniform grid spacing of $\Delta x/c = 0.02$. Each simulation is advanced for six convective time units, defined as $t=t^\prime U_\infty/c$, where $t^\prime$ denotes dimensional time. During the simulations, the following quantities are recorded for training and analysis: the vorticity field within a cropped subdomain $(-1.5c,3.3c) \times (-1.2c,1.2c)$, the surface pressure coefficient $\pres$ at 11 evenly-spaced prescribed sensor locations (see Fig.~\ref{fig:configuration}), the lift $\lift$ and drag $\drag$ coefficients (each defined as the respective force scaled by $\rho U_\infty^2 c/2$), and the instantaneous airfoil kinematics, including both the angle of attack $\ang$ and its angular velocity $\angv$. 

For computational efficiency and to ensure sufficient temporal separation between consecutive snapshots, data are stored every $\Delta t=3 \Delta t_s$, where $\Delta t_s$ denotes the numerical time step used in the solver. Because the maximum velocity varies across different randomly sampled cases, $\Delta t_s$---and consequently $\Delta t$---may differ from one simulation case to another, while remaining constant within each individual run. For instance, for a slowly pitching airfoil in randomly disturbed flow, $\Delta t \approx 0.024$. Accordingly, the total number of available snapshots varies across simulation cases. To maintain consistency during training---for the reasons discussed later---we retain the first 250 snapshots from each simulation. Although this results in slightly different final physical times among simulations, the retained window corresponds to approximately six convective time units in all cases. With a total of 100 simulated cases, this yields a dataset comprising $100 \times 250 = 25 \, 000$ data samples. The original state space to be estimated consists of the vorticity field, aerodynamic loads, and airfoil kinematics, while the available measurements include surface pressure coefficients and, for selected cases, localized vorticity slices introduced to enhance observability.

\subsection{Learning a reduced state space}\label{sec:learn_reduced_space}
\begin{figure*}
\centering
\includegraphics[width=1.0\textwidth]{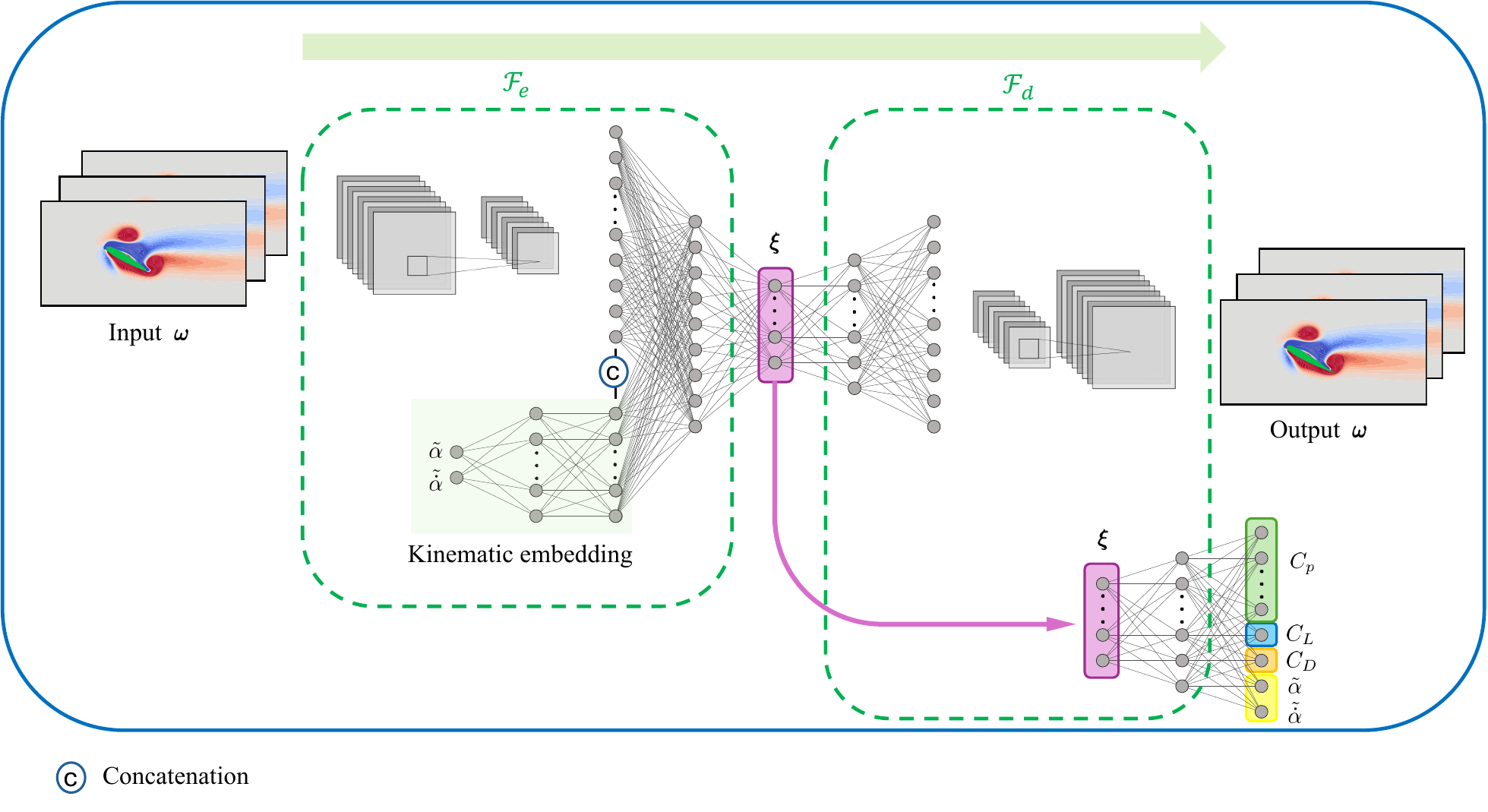}
\caption{\label{fig:network} Network architecture used in the present study. Convolutional encoder and decoder map the vorticity field to and from a low-dimensional latent representation. Kinematic variables are embedded using a multilayer perceptron and concatenated with vorticity features prior to projection into the bottleneck. From the latent vector $\lat$, additional multilayer perceptron heads predict aerodynamic loads, surface pressure coefficients, and airfoil kinematics.}
\end{figure*}

Having generated a sufficiently rich dataset for machine-learning purposes, the next essential step is to construct a low-dimensional representation of the high-dimensional aerodynamic state space. This reduced latent space, described by the coordinate vector $\lat$, serves as a compact surrogate for the vorticity field, aerodynamic loads, and motion kinematics associated with a pitching airfoil. Performing data assimilation in this reduced space is significantly more computationally efficient and numerically stable than operating directly in the original state space, whose dimension here is on the order of tens of thousands of degrees of freedom. We build upon the nonlinear autoencoder framework proposed by \citet{fukami2023grasping} for aerodynamic field compression. In particular, we extend the architecture developed in \cite{mousavi2025sequential}---originally formulated for a fixed airfoil in disturbed flow---to the more general setting of a pitching-up airfoil whose random kinematics are known only during training. The resulting architecture, which we will refer to as a \emph{kinematics-aware flow autoencoder}, is illustrated in Fig.~\ref{fig:network} with the detailed layer specifications summarized in Table~\ref{tab:network_blocks}.

As shown in Fig.~\ref{fig:network}, the network architecture for the pitching airfoil explicitly incorporates airfoil kinematics in both the encoder ($\encoder$) input and the decoder ($\decoder$) output. By embedding body kinematics within the latent representation learned through the autoencoder, the framework enables the estimator to implicitly disentangle flow variations induced by body motion from those arising due to flow disturbances, via multi-task reconstruction constraints. The high-dimensional vorticity field $\vor \in \mathbb{R}^l$ is first passed through a series of convolutional layers (Conv2D) to extract local spatial features, followed by max-pooling operations for downsampling. These features are then forwarded to a multi-layer perceptron (MLP) to capture global correlations. In parallel, the prescribed airfoil kinematics---namely the angle of attack and angular velocity, normalized to $\{\tilde{\ang}, \tilde{\angv}\} \in [-1,1]$---are processed through a separate MLP to extract kinematic features. These kinematic features are subsequently concatenated with the vorticity feature vector at the first fully connected layer of the encoder. The final layer of the encoder constitutes the bottleneck of the autoencoder and yields the latent representation $\lat \in \mathbb{R}^n$, where $n \ll l$. The decoder $\decoder$ then maps the latent vector back to the physical space, reconstructing the full set of flow variables and kinematic quantities,
\begin{equation}
    \{ \hat{\vor}, \h{\pmb{C}}_p, \h{C}_L, \h{C}_D, \hat{\tilde{\ang}}, \hat{\tilde{\angv}} \} = \decoder(\lat) = \decoder \left( \encoder(\vor, \tilde{\ang}, \tilde{\angv}) \right),
\end{equation}
where the $\hat{\cdot}$ denotes decoded (reconstructed) quantities.

The inclusion of both the airfoil angle and its angular velocity in the latent representation is a deliberate design choice. Conventional autoencoders lack explicit temporal information, and therefore may fail to distinguish states with nearly identical instantaneous flow fields but different kinematic histories. This issue is particularly pronounced for slowly pitching motions (small $\kramp$), where consecutive snapshots exhibit minimal changes in angle of attack but differ substantially in angular velocity and flow dynamics. In such cases, the encoder may otherwise map distinct physical states to the same point in the latent space. By explicitly incorporating angular velocity as an input and reconstruction target, the proposed kinematics-aware flow autoencoder ensures that the latent manifold remains sensitive to variations in pitching rate, thereby yielding a more expressive and dynamically informative reduced-order representation.

The network weights and biases $\weights$ are optimized by minimizing the following loss function with weight coefficients $\beta$:
\begin{align}\label{eq:loss}
    \weights = \underset{\weights}{\arg\min}(\mathcal{L}_{AE}) &= \underset{\weights}{\arg\min} \Big( 
    \underbrace{\beta_{\vor} ||\vor - \h{\vor}||_2^2}_{\text{vorticity reconstruction}} 
    + \underbrace{\beta_{\pres} ||\pres - \h{\pres}||_2^2}_{\text{pressure reconstruction}} \notag \\
    &+ \underbrace{\beta_{\lift} ||\lift - \h{C}_L||_2^2}_{\text{lift reconstruction}} 
    \ + \ \underbrace{\beta_{\drag} ||\drag - \h{C}_D||_2^2}_{\text{drag reconstruction}} \notag \\
    &+ \underbrace{\beta_{\ang} ||\tilde{\ang} - \h{\tilde{\ang}}||_2^2}_{\text{angle reconstruction}} 
    \ + \ \underbrace{\beta_{\angv} ||\tilde{\angv} - \h{\tilde{\angv}}||_2^2}_{\text{angular velocity reconstruction}}  \notag \\
    &+ \underbrace{\beta_{\lat} ||\lat - \mathcal{F}_{\text{inv}}(\h{\pmb{C}}_p)||_2^2}_{\text{latent observability regularization}}
    \ + \ \underbrace{\beta_{t} ||\lat_{t+1} - 2 \lat_{t} + \lat_{t-1}||_2^2}_{\text{temporal smoothness}} \Big),
\end{align}
where $\mathcal{F}_{\text{inv}}(\h{\pmb{C}}_p)$ is a lightweight auxiliary MLP network ($d \rightarrow 128 \rightarrow 128 \rightarrow n$) that maps the decoded pressure coefficients at $d=11$ sensor locations to the learned latent states $\lat$. This latent observability regularization is a simple additional term that encourages the learned latent variables to be predictable from pressure measurements alone, thereby promoting observability of the latent state through the available sensors. By constraining the latent variables to lie in directions that are sensitive to pressure, this regularization improves the conditioning of the latent-to-pressure mapping in the present problem, where the surface pressure reflects the coupled nonlinear effects of body motion, incoming disturbances, and unsteady flow dynamics. As a result, the latent manifold is regularized to align with the information content of the pressure observations, which improves robustness and stability during downstream data assimilation and filtering.

In general, the loss function defined in Eq.~\eqref{eq:loss} consists of three components: (i) a reconstruction loss for all decoder outputs, (ii) a latent observability regularization term, and (iii) a temporal smoothness loss. The temporal smoothness loss, which was also used in \citep{mousavi2025sequential}, is introduced to regulate the temporal evolution of the latent variables. This regularization enforces that $\lat(t)$ varies smoothly in time (approximately $C^2$ in the discrete sense), thereby suppressing spurious rapid fluctuations and improving the stability of downstream continuous-time modeling and sequential data assimilation \citep{mousavi2025sequential}. In this context, a fixed number of snapshots across simulation cases ensures a consistent temporal horizon during training, simplifies vectorized computation of the temporal loss, and prevents mixing of distinct flow cases within a single batch.

The loss weights $\beta_i$ in Eq.~\eqref{eq:loss}, with $i \in \{\vor, \pres, \lift, \drag, \ang, \angv, \lat, t \}$, are introduced to balance the relative contributions of the individual terms in the composite loss function. These weights are selected to emphasize accurate reconstruction of the vorticity and pressure fields, as the networks associated with these quantities serve as the primary observation operators in the subsequent data assimilation framework. Accordingly, the weights associated with the other loss components are chosen such that their weighted contributions remain at least one order of magnitude smaller than those of the vorticity and pressure reconstruction terms to speed up the convergence. Following empirical tuning, we adopt the set $\beta_{\vor} = \beta_{\pres} = \beta_{\lift} = \beta_{\drag} = \beta_{\ang} = \beta_{\angv} = \beta_{\lat} = 1$, together with $\beta_t = 500$ for the temporal smoothness regularization. It is important to emphasize that the comparatively large value of $\beta_t$ does not imply that the temporal smoothness term dominates the total loss. Rather, the discrete second-order temporal differences $||\lat_{t+1} - 2 \lat_{t} + \lat_{t-1}||_2^2$ are typically several orders of magnitude smaller than the reconstruction errors and therefore require a larger weighting factor to exert a comparable influence during optimization. (Since the temporal discretization is fixed throughout the dataset, any constant scaling associated with the finite-difference approximation is absorbed into the choice of $\beta_t$.) This scaling ensures that the temporal smoothness constraint is effectively enforced without overwhelming the reconstruction objectives.

The selected loss weights are not necessarily optimal in the sense of minimizing the overall reconstruction error. A comprehensive search over all weighting coefficients would be computationally prohibitive because of the size of the autoencoder architecture and the training dataset. Instead, the weights were selected using a goal-oriented strategy that places greater emphasis on accurately reconstructing the flow field and the surface pressure measurements, which are the primary quantities required for the subsequent state-estimation problem.
It should be emphasized that the learned latent variables only serve as an intermediate reduced-order representation for filtering. Consequently, different choices of the loss weights may produce different latent manifolds while yielding comparable reconstruction accuracy. As demonstrated in our previous work \citep{mousavi2025sequential}, sensitivity studies revealed that variations in these hyperparameters indeed modify the learned latent manifold. However, provided that the resulting latent space remains sufficiently observable through the available pressure measurements and the decoder reconstruction errors remain small, the subsequent estimation performance is largely unaffected. In this sense, the latent manifold acts primarily as a computational proxy for the filtering problem, and multiple latent representations may be equally suitable for accurate state estimation.

For training, $70\%$ of the independent random flow cases---each spanning the full duration of the coupled gust–pitching–airfoil interaction---are randomly selected for model training, while the remaining $30\%$ are reserved for validation and testing. This case-wise partitioning is deliberately adopted to reflect realistic deployment conditions, in which a model trained on a finite ensemble of stochastic aerodynamic realizations is applied to unseen flow scenarios. All network weights are initialized using Xavier (Glorot) initialization. This initialization scheme avoids degenerate initial latent trajectories that are nearly constant in time, which can otherwise arise when the encoder maps diverse flow snapshots to an overly compressed region of latent space during early optimization. Training is terminated once the validation loss converges and exhibits a sustained plateau (around epoch $\approx$ 4600), indicating that the model has reached a stable solution and further optimization yields negligible improvement or risks of overfitting.

\subsection{Sequential filtering in reduced space} \label{sec:filtering}
In the proposed framework, data assimilation is performed sequentially in the learned low-dimensional latent space introduced in Section~\ref{sec:learn_reduced_space}. As new sparse and noisy measurements $\y^*_k$ become available at time step $k$, we estimate the probability of the corresponding latent state $\lat_k$ in an online manner. Sequential filtering provides a principled Bayesian approach in which model predictions are recursively combined with streaming observations, yielding state estimates conditioned only on past and present observations in its exact Bayesian framework. This setting is particularly appropriate for transient aerodynamic flows, where measurements usually arrive sequentially, and online state estimation is desired. A comprehensive guideline on data assimilation in aerodynamic applications is provided by \citet{eldredge2025practical}.

Sequential filtering proceeds through two stages at each time step $k$: a \emph{forecast} (prediction) step and an \emph{analysis} (update) step. In the forecast step, the state distribution is propagated forward according to a dynamical model $\pmb{f}$. In fluid mechanics, $\pmb{f}$ may correspond to a high-fidelity CFD solver in the full space or a reduced-order dynamical model (ROM) that advances a projected state. When new measurements are acquired, the analysis step corrects the forecast distribution using an observation operator $\pmb{h}$ that maps the state to the observation space $\y$. The correction magnitude is governed by the innovation (the discrepancy between measured and predicted observations) together with state-observation cross-covariance and predicted observation covariance. 

In the learned latent space, the filtering objective is to approximate the posterior $\pi(\lat_k|\y_{1:k})$ from the prior $\pi(\lat_k|\y_{1:k-1})$ upon arrival of a new measurement $\y_k^*$. Conceptually, this operator transports probability mass from the forecast distribution to the analysis distribution. When the forecast and observation mappings are linear and all uncertainties are Gaussian, the posterior remains Gaussian and admits a closed-form Kalman filter update. In aerodynamic flow estimation, however, the dynamics and observation operators are typically strongly nonlinear---including when they are represented by data-driven surrogate models---so the induced state distributions are generally non-Gaussian. This motivates approximate nonlinear filtering methods; ensemble-based filters, as we use in this work, are particularly attractive due to their scalability: the EnKF represents the evolving state distribution using a finite ensemble and replaces exact covariance propagation with sample statistics, enabling efficient Bayesian updates in nonlinear and moderate- to high-dimensional settings \citep{eldredge2025practical, evensen2003ensemble, stuart2015data, le2021ensemble}.

Let $\Lat \coloneq \{\lat^i\}_{i=1}^M$ denote an ensemble of latent states. At each time step, the forecast model advances each ensemble member according to
\begin{equation}\label{eq:forecast}
    \lat^i_{k|k-1} = \frw(\lat^i_{k-1}) + \frwnoise^i_k,
\end{equation}
where $\frw(\cdot)$ is a learned latent dynamical operator and $\frwnoise_k^i \sim \mathcal{N}(\pmb{0},\pmb{Q})$ is additive process noise.
The forecast operator $\frw$ is learned as a Markovian one-step transition model in latent space using a Neural ODE architecture (Table~\ref{tab:network_blocks_forecast}). Training sequences are extracted which start a few snapshots after the latent trajectories start moving away from their undisturbed path, ensuring that the model is exposed to a diverse set of dynamically distinct initial conditions arising from different gust realizations. This strategy enhances the separability of latent trajectories associated with different disturbance parameters and improves the robustness of the learned dynamics model. To preserve the desirable Markovian characteristic of sequential filtering, we must ensure that the forecast provides consistent one-step prediction to propagate ensembles. At the same time, the learned dynamics must remain stable when rolled out over multiple steps to prevent ensemble divergence. To enforce both short-time accuracy and long-horizon stability, we train $\frw$ using a hybrid objective combining one-step and rollout losses as well as an equilibrium loss:
\begin{equation}
    \weights_{f} =  \underset{\weights_{f}}{\arg\min} \Big( \underbrace{ \beta_{f,1} ||\lat - \h{\lat}||_2^2}_{\text{rollout loss}} + \underbrace{\beta_{f,2}  ||\lat_{1:T} - \frw(\lat_{0:T-1})||_2^2}_{\text{one-step loss}} + \underbrace{\beta_{f,3}  ||\frw(\lat_{\text{eq}}) - \lat_{\text{eq}}||_2^2}_{\text{equilibrium loss}} \Big),
\end{equation}
where $\h{\lat}$ denotes the rolled-out prediction over the snapshot length of $T-1$ given the true initial condition. The last term is called the \emph{equilibrium loss} and it ensures that the learned forecast operator preserves equilibrium states---corresponding to a fixed airfoil in an otherwise undisturbed flow---by enforcing them as fixed points of the dynamics. We set $\beta_{f,1}=1$ and $\beta_{f,2}=\beta_{f,3}=100$. The selected weights prioritize accurate one-step propagation and equilibrium preservation, which are more important for the subsequent filtering problem than minimizing the open-loop rollout error alone, since the forecast model is repeatedly corrected through data assimilation. A comparatively small rollout weight is therefore sufficient to ensure that the propagated states remain stable and do not diverge over longer horizons.

When measurement $\y^*_k$ becomes available, we map each forecasted latent ensemble member to the observation space through
\begin{equation}\label{eq:observation}
    \y^i_k = \obs (\lat^i_{k|k-1}) + \obsnoise^i_k,
\end{equation}
where $\obs(\cdot)$ is the learned observation operator and $\obsnoise^i_k \sim \mathcal{N}(\pmb{0}, \obsCov)$ is i.i.d.\ observation noise with covariance $\obsCov$. In this work, $\obs$ is obtained directly from the pre-trained kinematics-aware flow autoencoder. For pressure-only measurements, $\obs$ corresponds to the pressure-decoding head $\h{\pmb{C}}_p = \obs(\lat)$. When the pressure observations are augmented with partial vorticity measurements, the observation operator is decomposed into two components. The first component corresponds to the pressure observations, $\h{\pmb{C}}_p = \obs^{(1)}(\lat)$, while the second corresponds to the vorticity observations obtained by applying a spatial restriction to the decoded vorticity field, $\mathcal{R}_{\mathcal{M}} \h{\vor} = \obs^{(2)}(\lat)$. Here $\mathcal{R}_{\mathcal{M}}$ denotes a restriction operator defined by the binary spatial mask $\mathcal{M}$, which extracts the entries of the decoded vorticity field corresponding to the measurement locations. The complete observation operator is then formed by concatenating these two components, $\obs(\lat) = \text{Concat}\left( \obs^{(1)}(\lat), \obs^{(2)}(\lat) \right)$.

In the vanilla stochastic EnKF (sEnKF), the analysis ensemble is obtained by transporting each forecasted particle through a data-dependent particle map,
\begin{equation}\label{eq:kalman_update}
    \lat_k^i = \lat_{k|k-1}^i + \gain \left(\y_k^* - \obs \! (\lat) - \obsnoise_k^i \right),
\end{equation}
with $\gain$ denoting the Kalman gain. In Eq.~\eqref{eq:kalman_update}, the quantity in the rightmost parentheses is the \emph{innovation}. To derive an expression for $\gain$, we first define the whitened variables $\tilde{\Lat} = \stateCov^{-1/2} (\Lat - \pmb{\mu}_{\Lat})$ and $\tilde{\Y} = \obsCov^{-1/2} (\Y - \pmb{\mu}_{\Y})$, where the observation ensemble is defined as $\Y \coloneq \{ \obs(\lat^i) \}_{i=1}^M$, with $\Lat \sim \mathcal{N}(\pmb{\mu}_{\Lat}, \stateCov)$ and $\Y \sim \mathcal{N}(\pmb{\mu}_{\Y}, \obsCov)$. This whitening standarizes the ensemble data to zero-mean and unit-covariance for simplicity. Then, in the sEnKF, $\gain$ can be derived following \citet{asch2016data} as 
\begin{equation}\label{eq:kalman_gain_senkf}
    \gain^{\text{sEnKF}} = \stateCov^{1/2} \stateObsCov \left(\obsObsCov + \pmb{I}_d \right)^{-1} \obsCov^{-1/2},
\end{equation}
with $\stateObsCov$ the sample cross-covariance between whitened latent-state and whitened observation ensemble matrices, and $\obsObsCov$ the covariance of the whitened observation ensemble. The collective action of Eq.~\eqref{eq:kalman_update} on the forecast ensemble induces a push-forward of the prior distribution to an ensemble approximation of the posterior. 

The matrix inversion appearing in the Kalman gain formula in Eq.~\eqref{eq:kalman_gain_senkf} is performed in the observation space and therefore involves a $d \times d$ matrix. While this is tractable for low-dimensional observations where $d \leq M$, the computational cost becomes prohibitive when the observation dimension $d$ is large. Since sequential filtering is intended for real-time estimation and feedback control, the analysis step must be carried out as efficiently as possible to avoid unacceptable latency. In ensemble-based filtering, when the observation dimension satisfies $d \gg M$, the ensemble-estimated observation covariance $\obsObsCov \in \mathbb{R}^{d \times d}$ is low rank, with rank at most $M-1$. This property enables a reformulation of the analysis step in ensemble space using the Woodbury identity, reducing the required matrix inversion to size $M \times M$ \citep{evensen2009data}. Define 
\begin{equation}
    \pmb{B} \coloneq \frac{1}{\sqrt{M-1}} \tilde{\Y} \in \mathbb{R}^{d \times M}.
\end{equation}
With this definition, we can conclude that $\obsObsCov = \pmb{B} \pmb{B}^{\top}$. Applying the Woodbury identity yields an equivalent analysis update in which only an $(M \times M)$ matrix is inverted. The resulting ensemble-space Kalman gain is given by
\begin{equation}\label{eq:kalman_gain_etkf}
    \gain^{\text{ES-EnKF}} = \stateCov^{1/2} \stateObsCov \left[ \pmb{I}_d - \pmb{B}\left( \pmb{I}_M + \pmb{B}^{\top} \pmb{B} \right)^{-1} \pmb{B}^{\top} \right] \obsCov^{-1/2}.
\end{equation}
This revised Kalman gain is then substituted in Eq.~\eqref{eq:kalman_update} to obtain the update formula corresponding to the ensemble-space EnKF (ES-EnKF) (similar to the ensemble transform Kalman filter \citep{bishop_adaptive_2001}). This formulation substantially reduces computational cost by restricting the dominant matrix inversion to the ensemble size $M$, enabling efficient filtering even for very high-dimensional observation spaces.

In this study, we switch between sEnKF (Eq.~\eqref{eq:kalman_gain_senkf}) and ES-EnKF (Eq.~\eqref{eq:kalman_gain_etkf}) depending on the relative magnitude of the observation dimension $d$ and the ensemble size $M$. Random gust disturbances in this study are introduced at arbitrary times with unknown strength, spatial extent, and orientation. Because the forecast operator has no prior knowledge of these incoming disturbances \citep{jones2022physics} and random body motions, it cannot predict the transition from an initially undisturbed flow to a disturbed state once a gust enters the domain or the airfoil starts maneuvering. Consequently, the estimator is designed to rely on frequent assimilation of measurements that capture the upstream signatures of approaching disturbances and body kinematics. 
The influence of assimilation frequency was thoroughly investigated by \citet{mousavi2025sequential} for a fixed-angle airfoil subjected to a strong transient gust. It was shown that assimilating measurements less frequently than every forecast interval, i.e., every $\Delta t$, fails to capture the rapid transient load spikes associated with gust encounter and results in substantially larger estimation errors. Accurate reconstruction of these fast dynamics therefore requires the assimilation of informative measurements at every forecast step. This requirement becomes even more critical in the present study, where the airfoil undergoes rapid pitch-up motion while simultaneously interacting with transient disturbances.
Accordingly, data assimilation is performed at the same temporal resolution as the learned dynamics, with the assimilation interval set equal to the dynamical time step $\Delta t$ for all cases. This configuration ensures that the estimator continuously incorporates new sensor information to detect incoming gusts and adapt to their varying strengths, orientations, and interactions with the pitching airfoil. Notably, the chosen assimilation frequency remains significantly lower than the typical sampling rates of pressure sensors. 

After performing sequential filtering in the reduced latent space, the flow quantities of interest are reconstructed by passing the estimated latent ensembles through the pre-trained decoder part of the kinematics-aware flow autoencoder. The estimation results are reported in terms of the ensemble mean together with twice the ensemble standard deviation, providing an estimate of the posterior mean and the associated $95 \%$ uncertainty bound for a Gaussian distribution. 
Throughout this study, the accuracy of vorticity estimation is quantified using the mean pixelwise relative $L_2$ error, defined as
\begin{equation}
    \varepsilon = ||\vor - \h{\vor}||_2/||\vor||_2.
\end{equation}
Before concluding this section on filtering, we highlight a key design choice in the proposed pipeline: the use of $\verb|Tanh|$ activation function throughout the encoder, decoder, and latent dynamics models. The rationale for this choice is discussed in Appendix~\ref{secap:activation}.

\section{Results and discussion}\label{sec:results}
This section presents and discusses the performance of the proposed framework for sequential estimation of unsteady flow fields, aerodynamic loads, and airfoil kinematics. Section~\ref{sec:estimation_pitching} demonstrates the capability of the framework to accurately estimate both body motion and flow states. Section~\ref{sec:shadowed_reconstruction} extends our previously proposed reduced-space sequential estimation framework for fixed-angle airfoils \citep{mousavi2025sequential} to the reconstruction of experimentally occluded regions.

\subsection{Estimating flow and airfoil kinematics}\label{sec:estimation_pitching}
This section addresses the primary contribution of the present study: the simultaneous estimation of the flow field, aerodynamic loads, and airfoil kinematics using sparse, noisy surface pressure measurements. The problem setup has been described in detail in Section~\ref{sec:problem}. Before presenting the estimation results for this fully coupled and highly unsteady configuration, it is instructive to first review the key flow physics associated with gust interactions with a pitching airfoil.

\subsubsection{Physics of the ramp pitch-up motion and gust encounters} \label{sec:physics}
Before presenting the state-estimation results for a pitching airfoil subjected to disturbances, it is instructive to briefly review the underlying flow physics associated with ramp pitch-up and hold motion and gust–airfoil interactions. The combined effects of unsteady airfoil kinematics and external vortical disturbances give rise to transient lift overshoots, circulation development, and complex vortex dynamics, all of which directly influence the aerodynamic loads and the surrounding flow field. Examining these mechanisms separately---for a pitching airfoil in a steady uniform flow, for a fixed airfoil encountering a gust, and for their combined interaction---provides essential context for interpreting the estimation results presented in the subsequent sections.

\begin{figure*}
\centering
\includegraphics[width=1.0\textwidth]{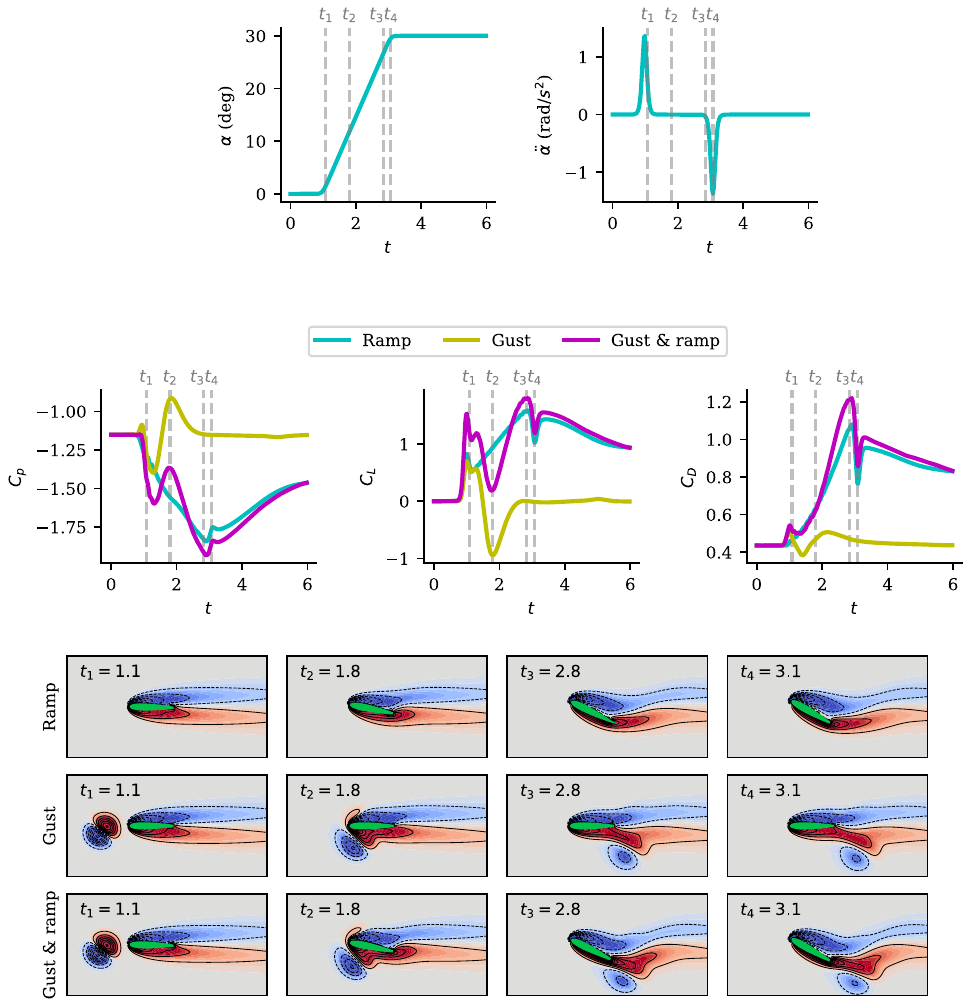}
\caption{\label{fig:pitching_ramp_gust_separation} Decomposition of aerodynamic response into distinct contributions: the gust acting on an airfoil fixed at zero angle of attack, the pitch-up motion of the airfoil in an undisturbed background flow, and the response of a ramping airfoil subjected to disturbed flow with the same gust parameters. The first row illustrates the pitch-up kinematics, the second row shows the corresponding aerodynamic loads, and the remaining panels depict the vorticity response. The pressure coefficient $C_p$ is plotted for sensor number 5 located on the suction side close to the LE. The conditions correspond to the case $\tramp=1.0$, $\kramp=0.125$, $D_y=-1.92$, $\sigma=0.15c$, $y_o=-0.12c$, and $t_o=1.0$.}
\end{figure*}

The kinematics of a representative random realization of the prescribed smooth ramp pitch-up motion are shown in the first row of Fig.~\ref{fig:pitching_ramp_gust_separation} (see the figure caption for the corresponding ramp parameters). For all cases considered, the ramp motion begins at approximately $t=1.0$, such that the airfoil remains at zero angle of attack in a steady uniform flow prior to the onset of pitching. Once the smooth ramp motion is initiated, the angle of attack increases approximately linearly in time (after the smoothed transition interval) at a randomly selected pitch rate (here $\dot{\ang} = 0.25$, or $\kramp = 0.125$), until reaching the prescribed maximum angle $\ang_{\max}=30^\circ$. To ensure smooth onset and termination of the motion, the ramp profile introduces nonzero angular acceleration and deceleration near the beginning and end of the maneuver, respectively, as evidenced by the temporal variation of $\ddot{\ang}$ shown in the figure.

For the discussion in this section, we disentangle the effects of gust interaction and prescribed ramp motion to examine their individual contributions before analyzing their combined influence on the flow field and aerodynamic loads at the end of this section. The corresponding results are shown in the middle row of Fig.~\ref{fig:pitching_ramp_gust_separation} for the aerodynamic loads and in the lower array of panels for the vorticity field. We first consider the prescribed pitch-up motion of the airfoil with kinematics illustrated in the first row of Fig.~\ref{fig:pitching_ramp_gust_separation} in an undisturbed uniform flow (cyan curves in the figure). The pressure coefficient $C_p$ shown for sensor~5, located on the suction side near the LE, remains negative throughout the motion, consistent with sustained suction at this location. As the airfoil undergoes a gradual pitch-up, the increasing angle of attack leads to increasingly negative suction pressure until approximately $t_3=2.8$, when the airfoil approaches the end of the ramp motion. Once the airfoil reaches its maximum pitch angle and the motion ceases ($t \geq t_4 = 3.1$), the suction at sensor~5 partially relaxes, reflected by a reduction in the magnitude of $C_p$. The corresponding lift coefficient $C_L$ exhibits an initial rapid increase from the onset of the pitch-up motion until $t_1=1.1$. Examination of the angular acceleration $\ddot{\ang}$ reveals that this early lift jump is primarily associated with added-mass effects arising from the nonzero angular acceleration of the airfoil at the start of the ramp motion. Following this initial transient, the lift increases approximately linearly with angle of attack, driven by the strengthening suction-side vorticity and enhanced trailing-edge (TE) shedding, as indicated by the extended regions of elevated vorticity in the contours at $t_2$ and $t_3$. This behavior is consistent with a progressive buildup of bound circulation. The lift reaches its maximum near $t_3=2.8$, after which it decreases slightly due to the added-mass response associated with deceleration as the ramp motion concludes. Beyond $t_4=3.1$, the lift levels off as the flow enters an ensuing periodic state (at later times not shown) at the final pitch angle $\ang_{\max}=30^\circ$. The drag coefficient exhibits a similar trend: it increases during the pitch-up phase as the wake thickens, peaks near $t_3$, and subsequently decays as the motion terminates and added-mass contributions diminish. These observations are consistent with those reported by \citet{granlund2013unsteady} at Reynolds number of $20 \, 000$.

Next, we consider the gust-only configuration, in which the airfoil is held fixed at zero angle of attack and subjected to a random vortical dipole gust introduced upstream. The initial state of the gust is visible in the vorticity field of the ``Gust'' row at $t_1=1.1$, corresponding to a gust introduced at $t_o=1.0$. In the load histories, the rapid initial variations in both the pressure coefficient $C_p$ and the lift coefficient $\lift$ prior to $t_1$ arise from non-circulatory effects associated with the velocity field induced by the newly introduced disturbance, even while the gust core remains upstream of the airfoil.
As the vortical gust convects downstream and reaches the LE ($t_2=1.8$), at such low Reynolds number---where viscous effects are important---its positive lobe interacts with the upper boundary layer \citep{jones2022physics}, distorting the near-body flow, as seen in the corresponding vorticity contour. This interaction weakens the local suction at sensor~5, producing a pronounced minimum in suction and a corresponding drop in lift at this instant. The flow response to this isolated gust is transient \citep{biler2021experimental}: the near-body vorticity rapidly reorganizes and is shed into the wake, leading to a quick recovery of both pressure and lift within approximately 1.7 convective time units (from $t_1$ to $t_3$). The absence of prolonged suction enhancement highlights that the gust alone does not produce sustained circulation buildup at a fixed angle of attack.
In contrast to the pronounced lift and pressure fluctuations, the drag response in the gust-only case is comparatively modest. A small transient increase in drag is observed during gust passage, associated with temporary wake deformation and enhanced vorticity shedding at the TE (see the vorticity fields at $t_2$ and $t_3$). However, once the gust convects downstream, the wake relaxes and the drag rapidly returns toward its pre-gust, undisturbed (and constant) value. By $t_4=3.1$, the gust has exited the near-field region, and the vorticity field approaches its pre-gust configuration.
Overall, variations in angle of attack have a more pronounced impact on drag than this specific Gaussian-distributed dipole gust disturbance at zero angle of attack. In contrast to the ramp-driven cases, the gust-only configuration does not exhibit a sustained drag increase, reflecting the absence of persistent circulation growth and kinematic forcing when the airfoil is held at a fixed angle of attack.

We now consider the combined case, in which the airfoil undergoes a prescribed ramp pitch-up motion while simultaneously interacting with a vortical gust (magenta curves and bottom row of vorticity snapshots in Fig.~\ref{fig:pitching_ramp_gust_separation}). For this case, both the ramp motion and the gust are introduced at the same time, $t=1.0$. In this configuration, the aerodynamic response reflects a nonlinear interaction between kinematic forcing and external disturbance.
At the onset of the motion, when the pitch-up acceleration is nonzero, both the ramp kinematics and the newly introduced gust contribute non-circulatory effects. During this initial phase, $t \in [1.0,1.1]$, these effects combine to produce an amplified transient lift response, shifting the early lift peak to a value at $t_1$ higher than that of the separate effect of pitch-up or gust. Following gust approach and its direct interaction with the airfoil, $t \in [t_1,t_3]$, the averaged evolution of lift, surface pressure, and drag in the combined case exhibits trends similar to those of the ramp-only configuration; during this interval, the gust response manifests as oscillatory modulations of the slowly varying lift and pressure evolution induced by the pitch-up motion.
The most pronounced deviations from the ramp-only behavior occur near $t_1$, due to the impulsive non-circulatory response associated with gust introduction, and near $t_2$, when the gust lobes interact directly with the airfoil and distort the near-field vorticity. Previous work by \citet{zaloglu2025} shows that the influence of a gust on the first lift peak depends on both the timing of gust encounter relative to motion onset and the speed of the pitch-up maneuver. As the gust convects downstream and approaches the trailing edge between $t_2$ and $t_3$, its direct influence on the near-field flow and aerodynamic loads diminishes. Consequently, the lift gradually relaxes toward values close to those of the ramp-only case as the gust-induced vorticity is shed into the wake and advected downstream. However, the peak lift at $t_3$ in the combined case exceeds that of the ramp-only case, as evidenced by the stronger suction observed in the $C_p$ signal. At this time, residual gust-induced vorticity in the near field modifies the local velocity distribution around the airfoil, effectively enhancing the instantaneous circulation. This effect is reflected in the vorticity contours by a stronger vorticity shed from the TE compared to the ramp-only configuration.
The drag response in the combined case is largely governed by the ramp motion but exhibits a systematic increase relative to the ramp-only configuration during gust–airfoil interaction, particularly over the interval $t \in [t_2,t_3]$. This increase reflects gust-induced thickening and distortion of the wake imposed on the wake development driven by pitch-up. The drag reaches its maximum near $t_3$, coinciding with the presence of a highly energetic, gust-modified wake. Once the gust has convected downstream and the pitch motion has ceased, both lift and drag settle toward their post-ramp values.

Overall, the combined gust–ramp case demonstrates that unsteady kinematics and external vortical disturbances interact nonlinearly to amplify transient aerodynamic loads and vorticity production. While the long-term response remains governed by the pitch-up motion, the gust significantly modulates the magnitude of load peaks by interacting with the developing viscous boundary layer and wake. This interaction highlights the importance of capturing both kinematic and environmental disturbances when modeling or estimating unsteady aerodynamic flows.

Having established physical insight into the respective roles of gust encounters, airfoil kinematics, and their interactions in shaping the aerodynamic loads and vortical flow structures, we now turn to the state-estimation problem.

\subsubsection{Estimation results using pressure sensors} \label{sec:estimation-pitching2}
For a pitching airfoil undergoing a prescribed ramp pitch-up and hold motion with randomly selected pitch rates and interacting with arbitrary vortical dipole gusts, Section~\ref{sec:physics} demonstrated that surface pressure measurements capture both circulatory and non-circulatory aerodynamic effects and therefore constitute informative observations for estimating the flow variables of interest. Accordingly, this section focuses on state estimation using 11 surface pressure sensors, denoted by $\pres$, whose locations on the airfoil are illustrated in Fig.~\ref{fig:configuration}.

Within the current estimation pipeline, the first step is to learn a compact, low-dimensional representation of the flow dynamics and motion kinematics. This is achieved using the kinematics-aware flow autoencoder introduced in Section~\ref{sec:learn_reduced_space}, which encodes the disturbed flow field, aerodynamic loads, and airfoil kinematics into a shared latent manifold. The resulting manifold serves as a reduced-order surrogate for the original high-dimensional aerodynamic-kinematic states. The dimensionality of the latent space, $\lat \in \mathbb{R}^n$, is prescribed prior to training. The appropriate choice of latent dimensionality depends on the complexity of the training loss function as well as on the number and nature of the output variables being reconstructed. To balance reconstruction accuracy and model compactness, we evaluate the performance of the autoencoder for different values of $n$ by monitoring the validation error. The reconstruction error is observed to saturate for latent dimensions larger than $n=10$. Consequently, all results presented hereafter employ a latent dimension of $n=10$, which provides an accurate representation of the system dynamics while maintaining a minimal latent dimensionality. Remarkably, the original state space---comprising tens of thousands of degrees of freedom---can be accurately represented using as few as ten latent dimensions.

In this section, surface pressure measurements are used as the sole observational input. Accordingly, the observation operator defined in Eq.~\eqref{eq:observation} corresponds to the pressure-decoder head of the learned kinematics-aware flow autoencoder, which maps latent states to pressure observations, i.e. $\obs: \mathbb{R}^{n=10} \rightarrow \mathbb{R}^{d=11}$. Since the measurement dimension is relatively small, we employ the stochastic EnKF (sEnKF) for the Kalman update defined in Eq.~\eqref{eq:kalman_update}, with the Kalman gain given by Eq.~\eqref{eq:kalman_gain_senkf}, to sequentially assimilate incoming streaming pressure measurements and approximate the posterior state distribution. The learned forecast operator introduced in Eq.~\eqref{eq:forecast}, with architecture summarized in Table~\ref{tab:network_blocks_forecast}, exhibits high accuracy for one-step latent-state predictions. The ensemble size is set to $M=200$ to adequately approximate the evolving distributions empirically. The initial latent ensemble is generated by perturbing the true initial latent state with a bias of $0.01$ and a variance of $10^{-4}$. The process-noise covariance $\pmb{Q}$ is determined empirically from the residual statistics between predicted and true latent trajectories over held-out test cases, thereby balancing model bias and ensemble spread. The measurement-noise variance is set to $10^{-4}$ for all pressure sensors. The influence of measurement noise on the estimation performance is systematically investigated in Appendix~\ref{secap:meas_noise}.

Remarkably, the complete set of 250 forecast–analysis cycles over six convective time units executes in less than one second on a single GPU, corresponding to approximately 4 milliseconds per cycle. This low computational cost demonstrates that the proposed estimator is highly efficient and can perform sequential updates with negligible computational overhead relative to the simulation timescale. Given the increasing integration of onboard GPUs in aerial platforms \citep{gao2018online, lin2018autonomous, loquercio2021learning}, these results suggest that the proposed reduced-order data-assimilation framework has strong potential for real-time aerodynamic state estimation in practical sensing and control applications.

The ensemble of estimated latent states is subsequently decoded using the operator $\decoder$ to obtain samples from the posterior distributions of the physically interpretable variables of interest, including the flow field and aerodynamic loads $\{\lift, \drag, \vor \}$, as well as the airfoil kinematics $\{\ang, \angv \}$. For reporting purposes, the reconstructed ensembles are approximated as Gaussian distributions and are summarized by their sample mean and standard deviation. Throughout the paper, the sample mean is denoted as ``Mean'' in the figures, while the $95 \%$ confidence interval is represented by twice the standard deviation and labeled as ``Uncertainty''. In the figures throughout this study, the vorticity fields labeled as ``true decoded'' correspond to reconstructions obtained by decoding the true latent trajectories through the trained autoencoder. This representation reflects the best achievable estimate within the reduced-order framework, as discrepancies between the decoded fields and the reference vorticity arise from decoder reconstruction error. All estimation results presented in this section---and throughout the manuscript---correspond exclusively to held-out test cases that were not used during training of any component of the learning framework.

\begin{figure*}
\centering
\includegraphics[width=1.0\textwidth]{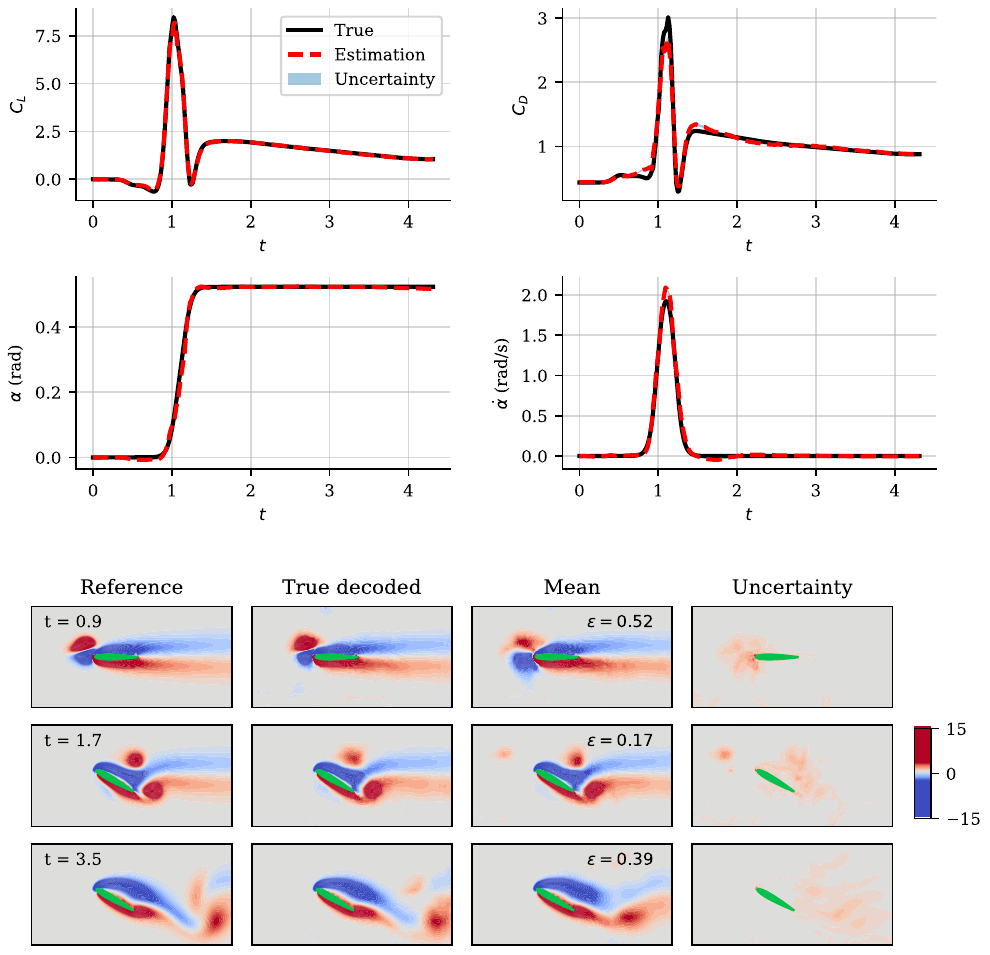}
\caption{\label{fig:pitching_case1} State estimation using surface pressure sensors for a pitching airfoil. The top panels report the reconstructed lift, drag, angle of attack, and angular velocity of a pitching airfoil, while the bottom panels present the vorticity field reconstruction at multiple time instances. Results are shown for the case $\tramp=1.0$, $\kramp=1.125$, $D_y=0.58$, $\sigma=0.08c$, $y_o=0.07c$, and $t_o=0.5$.}
\end{figure*}

Results for a representative case from the test set involving a random gust encounter with a pitch-up airfoil are shown in Fig.~\ref{fig:pitching_case1} (see the figure caption for the corresponding parameter values). The results demonstrate that the proposed estimator accurately reconstructs the aerodynamic loads and the prescribed airfoil kinematics. The posterior uncertainty bounds are generally narrow, indicating high confidence in the estimates; slight uncertainty is discernible in the drag prediction but remains small overall. The reconstructed vorticity fields provide additional insight into the relative timing of the pitch-up motion and gust–airfoil interaction. As indicated by the vorticity snapshots, the airfoil begins pitching at approximately $t \approx 1.0$, while the gust starts interacting significantly with the airfoil slightly earlier (see the vorticity field at $t = 0.9$). Although the decoder represents a highly nonlinear mapping from the latent space to the high-dimensional vorticity field, the ensemble-mean reconstructed vorticity closely matches the reference and true-decoded solution, particularly in the near-field region where the gust interacts with the pitching airfoil.
In contrast, the far wake region at later times (e.g., $t = 3.5$) is not reconstructed with the same accuracy. This region is only weakly observed by the 11 surface pressure sensors and therefore remains weakly constrained during assimilation. The limited observability of far-wake dynamics under surface-pressure-only measurements was analyzed in detail in our previous work \citep{mousavi2025sequential}, to which we refer the interested reader for a mathematical treatment. Consistent with these findings, the posterior uncertainty maps indicate relatively low uncertainty throughout most of the domain, with the highest uncertainty concentrated near the gust core, its immediate vicinity, and the shed vorticity. Similar estimation performance was observed across a range of gust cases with varying parameters and random timing relative to the pitch-up motion.

\begin{figure*}
\centering
\includegraphics[width=0.5\textwidth]{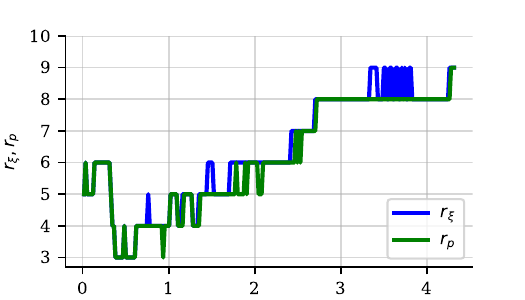}
\caption{\label{fig:pitching_lrenkf_ranks_case1} Effective state and observation ranks using surface pressure sensors during the update step of filtering for a pitching airfoil in disturbed flow. Results are shown for the case $\tramp=1.0$, $\kramp=1.125$, $D_y=0.58$, $\sigma=0.08c$, $y_o=0.07c$, and $t_o=0.5$.}
\end{figure*}

\begin{figure*}
\centering
\includegraphics[width=1.0\textwidth,trim={0 30 0 0},clip]{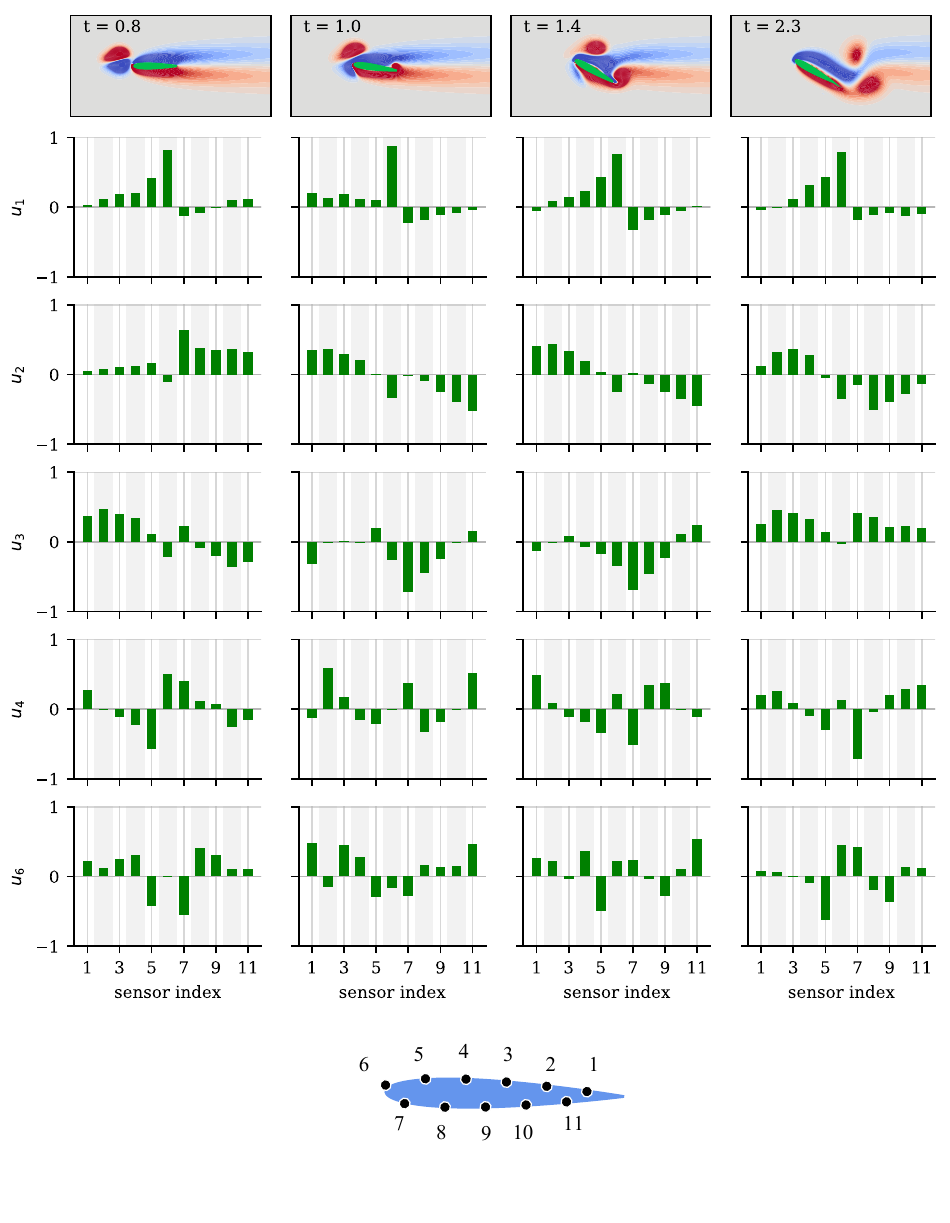}
\caption{\label{fig:pitching_lrenkf_obs_modes_case1} A few observation modes at different instants using surface pressure sensors as measurements for a pitching airfoil in disturbed flow. The $x$-axis corresponds to sensor index. Results are shown for the case $\tramp=1.0$, $\kramp=1.125$, $D_y=0.58$, $\sigma=0.08c$, $y_o=0.07c$, and $t_o=0.5$.}
\end{figure*}

The eigendecomposition of the state- and observation-space Gramians, as defined in \cite{mousavi2025sequential, le2022low}, reveals the dominant and trailing eigenvalues along with their corresponding eigenmodes. The effective state and observation ranks, $r_\xi$ and $r_p$, respectively, are determined by retaining $99 \%$ of the cumulative spectral energy of the aforementioned Gramians. For the disturbed pitching–airfoil case shown in Fig.~\ref{fig:pitching_case1}, the resulting rank evolution is presented in Fig.~\ref{fig:pitching_lrenkf_ranks_case1}, while the corresponding dominant observation modes are illustrated in Fig.~\ref{fig:pitching_lrenkf_obs_modes_case1}.

Figure~\ref{fig:pitching_lrenkf_ranks_case1} shows the history of effective state $r_{\xi}$ and observation ranks $r_p$ during gust encounter of a pitching-up airfoil. These ranks quantify the number of dynamically and observationally active directions (in the reduced space) at each assimilation time. Despite the relatively small dimensions of both the state and observation spaces, neither rank saturates at its theoretical maximum ($n=10$ for the state and $d=11$ for the observations). Instead, both $r_{\xi}$ and $r_p$ remain significantly lower and closely track each other throughout the assimilation window. 
In this example, a vortical dipole gust is introduced upstream at $t=0.5$, while the prescribed ramp pitch-up motion begins at $t=1.0$ and ends at $t=1.2$. Although the flow is initially fully developed and steady at zero angle of attack, the effective state and observation ranks exhibit a transient evolution during the early phase. This behavior is associated with filtering dynamics: the initial ensemble misalignment and forecast uncertainty briefly activate multiple latent directions, which are rapidly suppressed as surface pressure measurements collapse the ensemble toward the steady-state point. As a result, the effective ranks decrease to as low as 3 active modes.

Once the gust is introduced at $t=0.5$, the effective ranks increase from 3 to 4 as the disturbance convects toward the airfoil. The limited number of active modes during this phase indicates that the measurements are informative about the presence of the upstream disturbance, but not yet about its detailed interaction with the airfoil. This is consistent with the vorticity reconstruction shown in Fig.~\ref{fig:pitching_case1} at $t=0.9$, where the reconstruction error remains large for the predicted gust.

During the pitch-up interval, $t\in[1.0,1.2]$, the gust has already begun interacting with the airfoil, and the combined effects of kinematic forcing and external disturbance distribute the flow dynamics across additional latent directions. Accordingly, the effective ranks increase to 5 over this interval. After the airfoil reaches its maximum angle of attack, $\ang_{\max}=30^\circ$, the interacting disturbance and the developing wake further redistribute the spectral energy across a larger number of modes, leading to a gradual and sustained increase in both the effective state and observation ranks. 
This growth persists even as the core of the gust convects downstream into the wake, as seen in the vorticity field at $t=3.5$ in Fig.~\ref{fig:pitching_case1}. As the disturbance moves farther from the airfoil, the resulting flow structures increasingly occupy regions where the observation operator is less sensitive, and thus, where surface pressure measurements are less informative. Consequently, we have observed (without reporting) that the spectral energy of the Gramian spreads across additional modes required to represent the evolving wake dynamics. Ultimately, both the effective state and observation ranks saturate at nine.

The dominant observation modes shown in Fig.~\ref{fig:pitching_lrenkf_obs_modes_case1} provide further insight into the temporal evolution of information content in the surface pressure measurements. Each observation mode represents a spatially weighted linear combination of pressure sensors corresponding to an energetically important correction direction in latent space, and their evolution reveals how pressure sensitivity redistributes over the airfoil as the flow evolves.

\begin{figure*}
\centering
\includegraphics[width=1.0\textwidth]{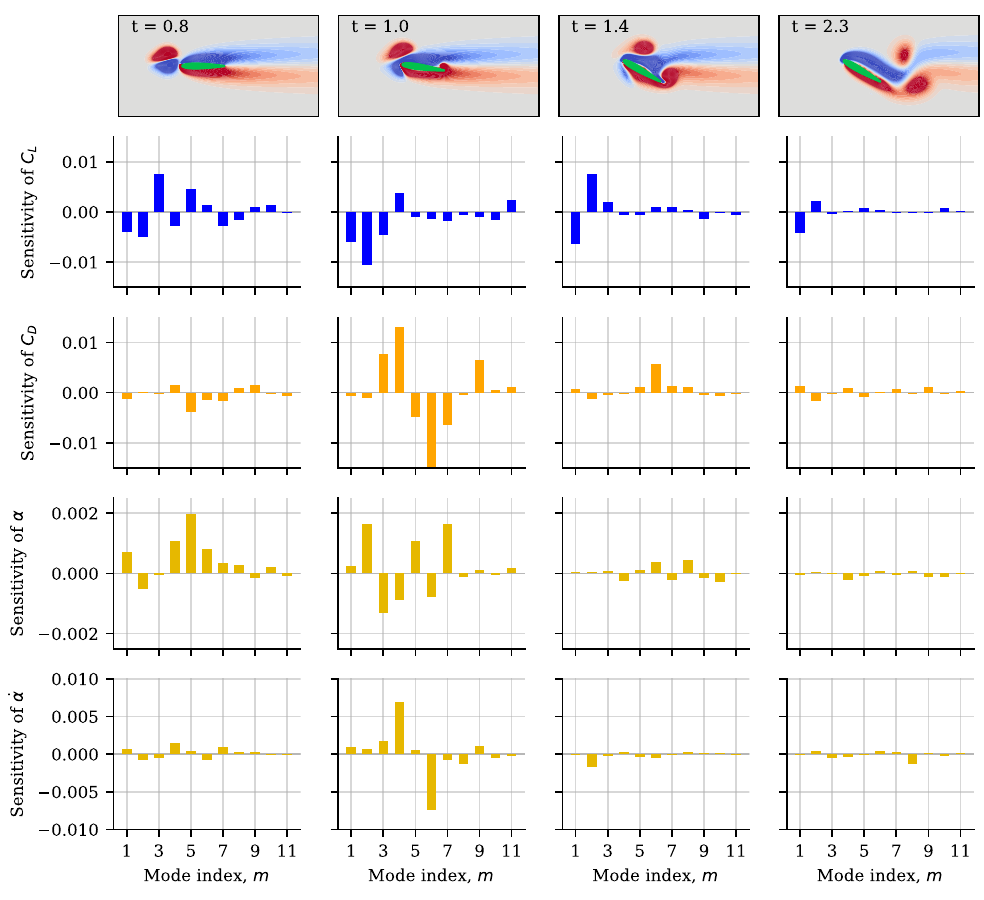}
\caption{\label{fig:pitching_lrenkf_variables_sensitivity} Sensitivity of output variables to unit perturbation in the direction of each measurement mode. Modes are ordered in decreasing contribution to the update step of filtering. The $x$-axis corresponds to observation mode index ordered by decreasing eigenvalues. Results are shown for the case $\tramp=1.0$, $\kramp=1.125$, $D_y=0.58$, $\sigma=0.08c$, $y_o=0.07c$, and $t_o=0.5$.}
\end{figure*}

Before examining the transient evolution of the observation-mode structures themselves, it is instructive to first quantify how a state variable of interest $D \in \{ \lift, \drag, \ang, \angv \}$ responds to pressure perturbations aligned with each observation eigen-direction. As illustrated by the kinematics-aware flow autoencoder architecture in Fig.~\ref{fig:network}, part of the decoder maps the latent state $\lat \in \mathbb{R}^n$ to both the pressure prediction $\pres(\lat) \in \mathbb{R}^d$ and the scalar quantity of interest $D(\lat) \in \mathbb{R}$. The sensitivity of $D$ to a unit perturbation along the $m$-th observation mode is defined as
\begin{equation} \label{eq:sensitivity}
    s^D_m = \mathbb{E}_{\lat \sim \pi(\lat)} \left[ \nabla D (\lat) \cdot \gain \cdot \pmb{u}_m \right],
\end{equation} 
where $\gain \in \mathbb{R}^{n \times d}$ denotes the Kalman gain, $\pmb{u}_m$ is the $m$-th observation eigenvector, and $\mathbb{E}_{\lat \sim \pi(\lat)} [\cdot]$ denotes the ensemble average. This quantity measures the linearized response of the analysis estimate of $D$ to perturbations along the corresponding observation eigen-direction.
The resulting sensitivities for the four quantities of interest are shown in Fig.~\ref{fig:pitching_lrenkf_variables_sensitivity} at four representative time instances during the pitching airfoil’s interaction with the traversing disturbance, corresponding to the snapshots shown in Fig.~\ref{fig:pitching_lrenkf_obs_modes_case1}.

Figure~\ref{fig:pitching_lrenkf_obs_modes_case1} demonstrates that at early times ($t=0.8$), when the gust is approaching the airfoil and nearing the LE, suction-side LE pressure variations define the single most energetic observation direction $u_1$, while pressure-side sensors contribute primarily to secondary, orthogonal correction directions captured by $u_2$.
Across all reported times, $t \in \{0.8, 1.0, 1.4, 2.3 \}$, the LE sensors (in particular, sensors 5 and 6) consistently carry significant weight in the dominant observation mode, indicating that pressure variations in this region remain the primary conduit through which the filter constrains the latent state during gust approach, interaction, and subsequent wake evolution. 
While the gust core's positive lobe moves over the suction side of the pitching airfoil and deforms the boundary layer there at $t=1.4$ and $2.3$, the suction side sensors dominate information in the first observation mode. 

The second dominant observation mode, $u_2$, exhibits an approximately antisymmetric structure about the LE sensor (sensor 6), with opposite signs on the suction and pressure sides at times $t \in \{1.0, 1.4, 2.3 \}$. A similar, though less pronounced, antisymmetric component is also present in the most energetic mode $u_1$ at these instants. Together, these structures indicate that the leading observation modes encode pressure differences across the airfoil, potentially consistent with the determination of lift. This interpretation is supported by the sensitivity analysis shown in Fig.~\ref{fig:pitching_lrenkf_variables_sensitivity}, where the lift coefficient exhibits pronounced sensitivity to observation modes~1 and~2 across the examined time instances. In contrast, the remaining quantities of interest---$\drag$, $\ang$, and $\angv$---display a broader sensitivity distribution, with substantial contributions from intermediate observation modes (modes~3–7) at all times. In particular, the drag response is not governed solely by the most dominant observation modes, but instead draws information from multiple measurement directions. This behavior is consistent with the fact that drag depends on both the distributed surface pressure and viscous shear stress along the airfoil, the latter of which is only indirectly observable via pressure variation along the wing surface, represented by intermediate-index pressure modes.
For the body kinematics, the sensitivities similarly reveal contributions from several observation modes, indicating mode mixing. This effect becomes more pronounced during periods of strong nonlinear interactions $t=0.8, 1.0$, reflecting the increased coupling between pressure distributions and kinematic responses under highly unsteady conditions. At $t=1.0$, when the body has just started pitching, angular velocity exhibits its strongest sensitivity to perturbations aligned with the $4^{\text{th}}$ and $6^{\text{th}}$ observation modes. Inspection of the corresponding sensor weight distributions in Fig.~\ref{fig:pitching_lrenkf_obs_modes_case1} shows that these modes place significant emphasis on sensors located near the TE. This behavior is physically consistent with the pitching kinematics: for an airfoil pitching about the quarter-chord point, the TE experiences the largest translational velocity due to its maximum distance from the pivot point. As a result, pressure measurements near the TE are most informative of instantaneous angular velocity, explaining the dominant sensitivity of $\angv$ to TE sensors during the pitch-up maneuver.

For all quantities of interest, the mode-wise sensitivities peak during the gust's approach, its interaction with the airfoil, and the pitch-up phase of the body motion, up to $t=1.2$ when the airfoil reaches its maximum angle of attack and the motion is arrested. During this interval, the learned forecast operator alone is incapable of detecting the unknown approaching disturbance and motion kinematics. In fact, surface pressure measurements play a dominant role in early times in feeling the flow changes and correcting the state estimate accordingly, leading to pronounced sensitivities across the multiple observation modes. Once the airfoil reaches and maintains its maximum angle of attack and the disturbance convects downstream, the estimation enters a regime in which the state variables become effectively locked onto their forecasted trajectories. In this regime, the Kalman gain diminishes and the analysis updates become progressively weaker.

As mentioned before, Fig.~\ref{fig:pitching_lrenkf_variables_sensitivity} demonstrates the non-negligible contribution of intermediate observation modes to the estimation of $\drag$, $\ang$, and $\angv$ at early times. As shown by the rank histories in Fig.~\ref{fig:pitching_lrenkf_ranks_case1}, by $t=1.2$ the effective rank corresponds to the first five dominant observation eigenmodes. Nevertheless, modes~6 and~7 continue to exhibit nontrivial importance for $\drag$, $\ang$, and $\angv$. By definition, these higher-index (lower-energy) modes are weakly observable in the latent–pressure update. However, the latent directions informed by these modes can still play an important role in determining the decoded outputs. As a result, aggressive rank truncation at early times can degrade the estimation accuracy of certain variables.

\begin{figure*}
\centering
\includegraphics[width=1.0\textwidth,trim={0 0 0 0},clip]{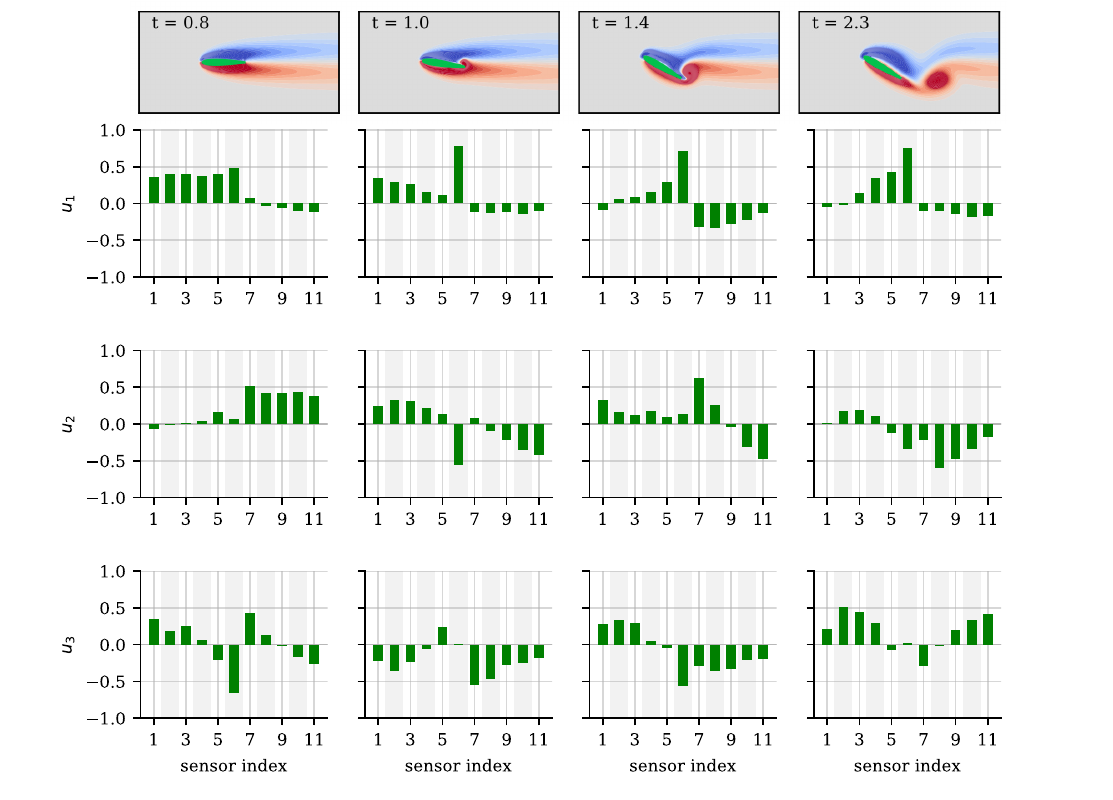}
\caption{\label{fig:pitching_lrenkf_obs_modes_pitching_only} The first three dominant observation modes at different instants using surface pressure sensors as measurements for a pitching airfoil in the undisturbed background flow. Results are shown for a pitching airfoil with $\tramp=1.0$, $\kramp=1.125$ (similar to Fig.~\ref{fig:pitching_lrenkf_obs_modes_case1} but without the gust).}
\end{figure*}

To isolate the effect of the body motion on the structure of the observation modes, we next consider the same pitch-up motion as in the previous case, but now in an otherwise undisturbed flow. In this pitch-up–only configuration, all flow unsteadiness arises solely from airfoil kinematics. The corresponding results are shown in Fig.~\ref{fig:pitching_lrenkf_obs_modes_pitching_only}. 
The first three dominant observation modes are shown at the same time instants as in Fig.~\ref{fig:pitching_lrenkf_obs_modes_case1} to facilitate direct comparison. At the early time $t=0.8$, when the airfoil remains at zero angle of attack in a steady flow and has not yet begun pitching, the dominant observation mode differs noticeably between the disturbed and undisturbed cases. In particular, the LE sensor (sensor 6) carries substantially greater weight in the disturbed case, reflecting its heightened sensitivity to the change in the effective angle of attack induced by an approaching gust \citep{fukami2023grasping}. Once the pitch-up motion begins at $t \approx 1.0$, the LE sensor (sensor 6) becomes the most informative sensor in both cases and remains dominant thereafter. This behavior reflects the central role of LE pressure in sensing changes in the angle of attack associated with pitching kinematics.

The second dominant mode $u_2$ exhibits a similar structure in both cases at pitch-up onset ($t=1.0$), corresponding to the initial non-circulatory lift response associated with the onset of angular acceleration. At this stage, the flow remains largely governed by kinematics, and gust-induced disruption of the near-body vorticity field is not yet significant. At later times ($t=1.4$), however, the presence of a gust in the disturbed case and its interaction with the near-field vorticity redistributes the sensor weights in $u_2$. Interestingly, for the pitching airfoil in either the undisturbed or disturbed case, the TE sensor becomes increasingly informative in the second dominant observation mode $u_2$ during the pitch-up phase and shortly after the airfoil reaches its maximum prescribed angle. This elevated importance is associated with the higher flow velocity close to the TE and reflects the sensitivity of the TE pressure to the growth of the TE vortex and the progressive development of the wake during the pitch-up motion. Once the first TE vortex is shed---shortly after $t \approx 1.4$---the flow near the TE changes less dramatically, and the TE sensor correspondingly becomes less informative in subsequent observation modes.

Finally, at $t=2.3$, once the gust in the disturbed case has convected sufficiently downstream and away from the airfoil, its influence on the surface pressure measurements diminishes. At this time, the dominant observation modes closely resemble those of the pitch-up–only case, indicating that the near-body flow is once again governed primarily by the airfoil at its maximum angle of attack. This convergence demonstrates that gust-induced modifications to the observation modes are transient and localized in time, while the long-term structure of the informative subspace is dictated by the pitching motion and the overall change of airfoil configuration.

\subsubsection{Improving observability}\label{sec:slice_measurements}
Our prior work \citep{mousavi2025sequential} and the results of the previous section demonstrated that, for disturbed flow over an airfoil, estimating high-dimensional flow states from sparse surface pressure measurements via filtering in the latent space can be limited by partial observability. In particular, when the measurement vector consists solely of surface pressure sensors, certain regions of the flow field remain weakly informed by the observations. These poorly observable regions were found to occur primarily (i) in the upstream region, and (ii) in portions of the downstream field where pressure signatures become weak. An example of such a weakly-observed region for the disturbed pitching–airfoil case is illustrated in Fig.~\ref{fig:pitching_case1}, where large estimation errors appear in the far wake at $t=3.5$. In \citet{mousavi2025sequential}, these regions were characterized more formally by perturbing the latent state along directions in the null space of the state-space Gramian (a quantity based on the Jacobian of the observation operator, $\nabla \obs$). While pressure measurements alone have nonetheless yielded accurate estimates around the airfoil, it is still useful to understand how extending the physical measurements to other locations and quantities might improve the observability, e.g., for early detection of the timing and strength of incoming disturbances.

To that end, we consider idealized measurements of the out-of-plane vorticity field, $\vor(x,y)$, sampled along a small number of vertical slices. Along each slice, vorticity is assumed to be measured at every grid point in the $y$-direction, such that the number of vorticity sensors per slice equals the number of grid points in the vertical direction (here 120). This relatively dense sensor arrangement ensures that random incoming disturbances intersect at least one sensor along each slice. 
Vorticity provides a compact diagnostic of vortex dynamics and is therefore convenient for probing information content and guiding sensor placement. In this section, slice measurements are appended to the pressure observations to form an augmented measurement vector. We emphasize that vorticity, while easily accessible in two-dimensional simulations, is not typically measurable directly in most operational settings. Nevertheless, \citet{alsalman2018training} showed that the transverse velocity component can contain nearly the same information content as vorticity for wake inference in two-dimensional flows, with vortex passage signatures captured strongly in this transverse velocity component. Consequently, the observability trends inferred here using $\vor$-slice measurements can be interpreted as an upper bound on the information that would be available from more practical realizable transverse-velocity measurements, and the identified slice locations may be viewed as proxies for effective velocity measurements.

\begin{figure*}
\centering
\includegraphics[width=1.0\textwidth]{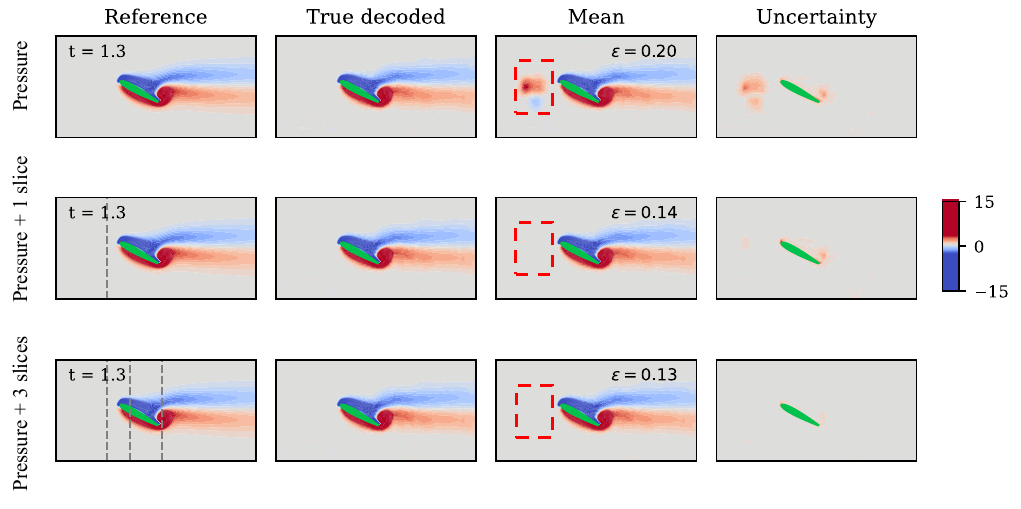}
\caption{\label{fig:pitching_slices_case3_t_09} Observability improvement using surface pressure and vorticity slice measurements for a pitching airfoil in undisturbed flow at $t=1.3$. The reported uncertainty interval corresponds to $95 \%$ uncertainty bound. Results are shown for the case $\tramp=1.0$, $\kramp=1.0$, $D_y=0.26$, $\sigma=0.1c$, $y_o=-0.23c$, and $t_o=1.8$.}
\end{figure*}

Accordingly, in addition to the 11 evenly-spaced surface pressure sensors shown in Fig.~\ref{fig:configuration}, we consider two sets of vorticity-slice configurations motivated by the observability deficiencies reported earlier. First, to target the loss of upstream observability prior to gust impact, we place a single vertical slice at $x_{s,1}=-0.3$ (with the LE located at the origin). Second, we consider three slices intended to inform the upstream approach region, the near-airfoil interaction region, and the downstream wake, respectively, at $x_{s,1}=-0.3$, $x_{s,2}=0.5$ and $x_{s,3}=1.1$. The relative locations of these slices and the three corresponding measurement sets—--(i) pressure only, (ii) pressure plus one upstream slice, and (iii) pressure plus three slices—--are illustrated in the left column of vorticity in Fig.~\ref{fig:pitching_slices_case3_t_09} for a pitching-up airfoil in a disturbed flow.

An important practical feature of this setup is that augmenting the measurement vector \textbf{does not require retraining} the learned models once the convolutional autoencoder has been trained, since all measurement modalities considered here are already available among the decoder networks in the original architecture. Additional measurement types are incorporated into data assimilation by expanding the definition of the observation operator with the appropriate decoder heads. For vorticity slices, we apply a spatial restriction (masking) operation to the vorticity decoder output, as described following Eq.~\eqref{eq:observation}. This allows the observation operator to be modified online to reflect arbitrary combinations of decoder outputs (e.g., pressure, partial vorticity, loads) while keeping the encoder/decoder/NeuralODE weights fixed.

Since the dimension of the observation vector still remains relatively small and comparable to the ensemble size, the sEnKF can be employed to efficiently estimate the latent states. For sEnKF inference, the setup follows that of Section~\ref{sec:estimation-pitching2}. Measurement-noise variances are set to $10^{-4}$ for surface pressure measurements and $10^{-1}$ for vorticity-slice measurements; these values are chosen to be commensurate with the reconstruction errors of the decoder, noting that the vorticity reconstruction error is typically larger than that of surface pressure. With these settings, assimilating 250 cycles requires only a few seconds of wall-clock time, demonstrating the computational efficiency of reduced-space filtering.

\begin{figure*}
\centering
\includegraphics[width=1.0\textwidth]{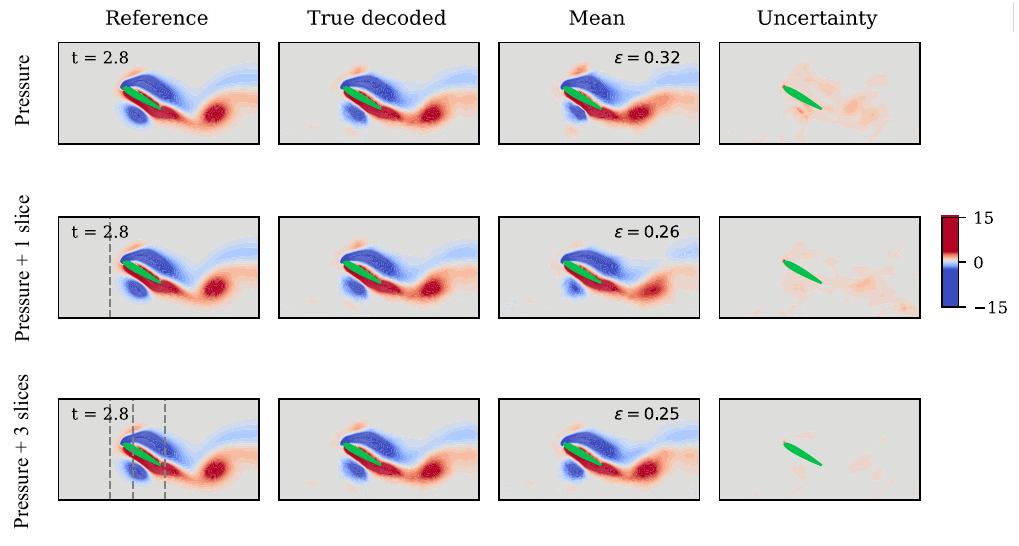}
\caption{\label{fig:pitching_slices_case3_t_26} Observability improvement using surface pressure and vorticity slice measurements for a pitching airfoil in disturbed flow at $t=2.8$. The reported uncertainty interval corresponds to $95 \%$ uncertainty bound. Results are shown for the case $\tramp=1.0$, $\kramp=1.0$, $D_y=0.26$, $\sigma=0.1c$, $y_o=-0.23c$, and $t_o=1.8$.}
\end{figure*}

\begin{figure*}
\centering
\includegraphics[width=1.0\textwidth]{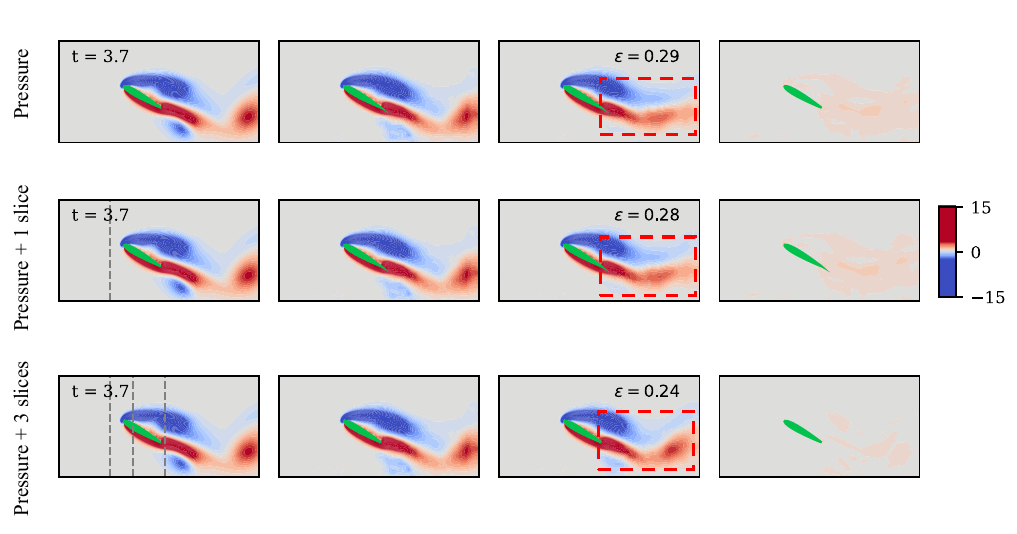}
\caption{\label{fig:pitching_slices_case3_t_37} Observability improvement using surface pressure and vorticity slice measurements for a pitching airfoil in disturbed flow at $t=3.7$. The reported uncertainty interval corresponds to $95 \%$ uncertainty bound. Results are shown for the case $\tramp=1.0$, $\kramp=1.0$, $D_y=0.26$, $\sigma=0.1c$, $y_o=-0.23c$, and $t_o=1.8$.}
\end{figure*}

The estimation results for the three measurement configurations, obtained for a different random realization of the disturbance and motion kinematics, are shown in Figs.~\ref{fig:pitching_slices_case3_t_09}, \ref{fig:pitching_slices_case3_t_26}, and \ref{fig:pitching_slices_case3_t_37} at three representative time instances. This particular realization corresponds to a case in which the pressure-only estimation experienced a strong upstream hallucination at early times, and is therefore selected as a representative worst-case example for assessing the improvement in observability provided by the augmented measurement configurations. The result of the estimated loads and motion kinematics is omitted in this section to avoid redundancy. As observed earlier, surface pressure measurements alone already provide accurate estimates of these quantities, and augmenting the measurement vector with additional observations has only a negligible effect on their estimation accuracy. 
The vorticity fields shown in Fig.~\ref{fig:pitching_slices_case3_t_09} indicate that surface pressure sensors alone exhibit limited sensitivity to upstream flow variations. As a result, at $t=1.3$, the estimator exhibits a hallucination, incorrectly inferring the presence of a dipole gust approaching the airfoil, manifested in both the reconstructed vorticity mean and in the elevated uncertainty concentrated near the reconstructed gust core, despite the fact that no gust is actually present in the reference domain at this time (for this case, the gust is introduced at $t_o=1.8$; see figure caption). This spurious upstream structure is highlighted by the red dashed box in the estimated mean vorticity field and reflects a significant discrepancy between the reconstructed and reference solutions. When a single upstream slice of vorticity measurements is added to the measurement vector, both the hallucinated gust and its associated uncertainty within the highlighted region are substantially reduced. This improvement arises because the upstream vertical arrangement of vorticity sensors lies within the weakly observed region and directly constrains the latent states through the assimilation of local vorticity information. Incorporating two additional vorticity slices, however, has minimal influence on the upstream estimation and uncertainty, and mainly reduces the estimation error in regions closer to the mid-chord and wake.  

When the airfoil has already ceased pitching and the gust is introduced near the left boundary at $t_o=1.8$ and subsequently convects downstream to interact with the airfoil at $t=2.8$, Fig.~\ref{fig:pitching_slices_case3_t_26} reveals a pronounced estimation error in the gust core, the deformed upper boundary layer, and the positive shed vorticity from the TE when only surface pressure measurements are assimilated. These flow features are poorly reconstructed under the pressure-only configuration, but are more directly informed in the configurations that augment surface pressure measurements with vorticity slices, as shown in the bottom two rows of Fig.~\ref{fig:pitching_slices_case3_t_26}.  

This interpretation is further supported by the later-time results shown in Fig.~\ref{fig:pitching_slices_case3_t_37}. At $t=3.7$, when the gust has convected far downstream and the TE shed vortex is exiting the domain, both the surface-pressure-only configuration and the configuration augmented with a single upstream vorticity slice erroneously predict a diffused wake with no disturbance or shed vorticity within the red dashed box. The largest uncertainty levels in these cases are concentrated within the highlighted region of the wake. By comparison, when three vorticity slices are incorporated into the measurement vector, the downstream wake slice accurately captures both the TE shed vortex and the intensified negative lobe of the gust at their correct locations. Consequently, the reconstructed wake closely resembles the true-decoded vorticity field, with a noticeable reduction in the mean reconstruction error. In this case, continuous assimilation of vorticity along the downstream-most slice and the subsequent convection of vorticity by the background flow ensure that downstream flow features are persistently corrected as they convect, resulting in reduced estimation error and significantly lower uncertainty in the far-wake region.

Across all time instances, and consistent with the qualitative observations discussed above, both the estimation error $\varepsilon$ and the maximum posterior uncertainty decrease monotonically as the measurement configuration is enriched—from surface-pressure-only measurements, to pressure augmented with a single vorticity slice, and finally to pressure combined with three vorticity slices. We observed (not reported here for brevity) that using vorticity slice measurements alone, without surface pressure data, leads to poor estimation performance, as each slice provides only localized flow information. This reflects the critical role of surface pressure sensors in constraining the flow dynamics in the vicinity of the airfoil, particularly during gust–airfoil interactions where accurate near-field reconstruction is essential.
Together, these results highlight the complementary information provided by the different measurement types. As expected, surface pressure measurements primarily provide global information about the flow, with their influence gradually diminishing with distance from the airfoil, while also directly informing the aerodynamic loading. Additionally, the upstream vorticity measurements enable early detection and characterization of incoming gusts, and downstream vorticity slices constrain the wake evolution by continuously correcting the flow as it convects. The combined use of surface pressure sensors and strategically placed vorticity slices, therefore, yields the most accurate and robust flow-state estimation, clarifying the distinct and synergistic roles that each measurement modality plays within the sequential filtering framework.

\subsection{Flow completion for experimentally missing data}\label{sec:shadowed_reconstruction}
Beyond estimating body kinematics and aerodynamic states during gust encounters, the estimation framework can also be used as a means of reconstructing experimentally missing data. This section focuses on recovering flow fields within shadowed regions commonly encountered in laboratory aero(hydro)dynamic image-based measurements, such as with particle image velocimetry. Accurate estimation of the flow dynamics in these unobserved areas reveals the mechanisms by which incoming gust disturbances interact with the airfoil and shape the ensuing wake evolution, thereby providing predictive insight that extends beyond what can be inferred from surface pressure measurements alone. In this section, we assess the ability of the reduced-space sequential filtering framework, applied to the fixed-angle airfoil configuration introduced in our previous work \citep{mousavi2025sequential}---where data was compressed into seven-dimensional latent a space--- to reconstruct flow fields in the presence of data gaps. It should be noted that the framework developed for a fixed-angle airfoil is consistent with the pitching-airfoil framework presented in this study, except that motion kinematics are not included in the convolutional autoencoder.

\begin{figure*}
\centering
\includegraphics[width=1.0\textwidth]{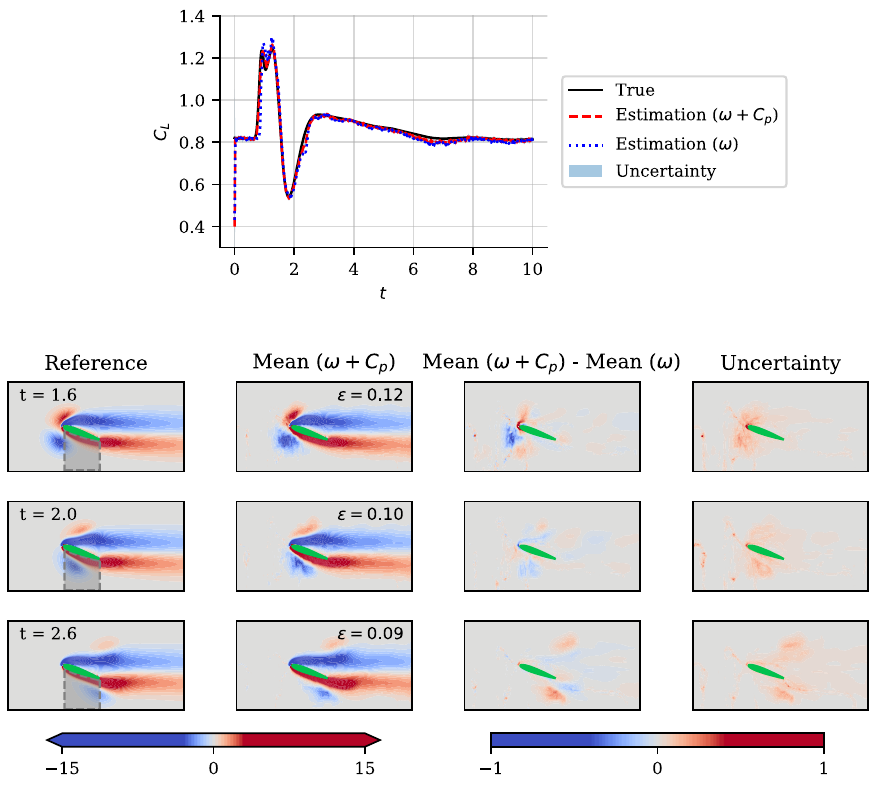}
\caption{\label{fig:AoA20_whole_plane} State estimation using surface pressure measurements and planar vorticity data outside the shadowed region beneath the airfoil. The top panel shows the reconstructed lift history, while the bottom panel presents the reconstructed vorticity fields at selected time instances. Results are compared against a baseline case that uses only planar vorticity measurements outside the shadowed region (denoted by $\vor$), highlighting the impact of incorporating surface pressure data (denoted by $(\vor+\pres)$) on flow-state estimation in this configuration. The reported uncertainty interval corresponds to $95 \%$ uncertainty bound. Results are shown for the case $\alpha=20^\circ$, $D_y=-0.58$, $\sigma=0.18c$, $y_o=-0.16c$, and $t_o=0.9$.}
\end{figure*}

To this end, we consider a measurement configuration in which surface pressure measurements are augmented with planar vorticity measurements over the entire two-dimensional flow field, except within a shadowed region beneath the airfoil. The out-of-plane vorticity field, \(\vor\), is usually computed from planar PIV measurements of the in-plane velocity components \(u\) and \(v\). An example of the resulting shadow geometry---corresponding to a laser sheet introduced from above the airfoil and illuminating the suction side---is shown in the left column of the vorticity fields in Fig.~\ref{fig:AoA20_whole_plane} for an airfoil at \(\ang = 20^\circ\).

To emulate dense experimental measurements, we assume one vorticity sensor at each computational grid point outside the shadowed region. We compare estimation results obtained using combined surface pressure and vorticity measurements (denoted by \(\vor + \pres\) in the figures) with those obtained using vorticity measurements alone (denoted by \(\vor\)), in order to assess the role of surface pressure information when the vorticity observation space is already high-dimensional. In particular, this setup allows us to examine whether the estimator can detect and reconstruct a gust---or individual gust lobes---passing through the shadowed region, even though the corresponding vorticity is not directly measured.

Because the observation dimension \(d\) in this configuration is extremely large (on the order of $2 \times 10^4$ in the present study) and satisfies \(d \gg M\), where \(M = 200\) is the ensemble size, we employ the ES-EnKF formulation to reduce computational cost during the analysis step. For inference, the latent ensemble \(\lat\) is initialized by offsetting the true initial latent state by 0.5 and sampling \(M = 200\) ensemble members around this biased mean with variance 0.25. The process-noise variance is set to \(10^{-2}\). Measurement-noise variances are set to \(10^{-4}\) for surface pressure measurements (when included) and \(10^{-1}\) for vorticity measurements, similar to the setup in Section \ref{sec:slice_measurements}. Under these settings, assimilating 500 snapshots corresponding to 10 convective time units requires only a few seconds of wall-clock time, demonstrating the computational efficiency and scalability of reduced-space filtering using the ES-EnKF in the presence of high-dimensional observations.

The estimation results are shown in Fig.~\ref{fig:AoA20_whole_plane} for an airfoil at \(\ang = 20^\circ\). The true lift exhibits a sharp transient peak immediately after the vortical gust is introduced upstream at $t=0.9$. This initial response is dominated by non-circulatory effects, which arise from the rapid adjustment of the pressure field required to enforce the no-penetration boundary condition as the gust-induced velocity impinges on the airfoil surface. As the gust convects closer to the LE, the lift response transitions to being governed primarily by circulatory effects, associated with the establishment and evolution of bound circulation and wake vorticity, leading to comparatively slower variations in lift for $t \gtrsim 1.0$. 

The estimation results further demonstrate that the filter is largely insensitive to the choice of the initial ensemble distribution, as the reconstructed states converge to the true states within only a few assimilation cycles. Figure~\ref{fig:AoA20_whole_plane} highlights the distinct role of surface pressure measurements in estimating both lift and the vorticity field. When surface pressure and vorticity measurements are assimilated jointly (red dashed curve, $\vor + \pres$), the estimated lift closely tracks the true lift throughout the gust encounter, accurately capturing both the initial rapid rise and the subsequent transient peak. This behavior is expected, since lift is physically obtained from the surface pressure integral, and surface pressure measurements directly encode both the fast non-circulatory response and the slower circulatory contribution. In contrast, when only vorticity measurements are assimilated (blue dotted curve, $\vor$), the estimated lift exhibits a phase lag during the early stage of the gust approach, despite the gust vorticity being fully observed outside the shadowed region. This lag appears immediately after gust introduction and persists until approximately $t=1.0$, when circulatory effects start to dominate. The observed delay arises because the non-circulatory response manifests in the vorticity space primarily through boundary-layer vorticity, as shown by \citet{corkery2019quantification}. However, a portion of this boundary-layer region is obscured within the shadowed area beneath the airfoil, leading to incomplete observability of the non-circulatory contribution from vorticity measurements alone. The inclusion of surface pressure measurements compensates for this missing boundary-layer information. Consequently, the vorticity-only estimate underpredicts the early, impulse-like lift response and converges rapidly to the correct lift evolution only once the gust–airfoil interaction produces sufficiently strong and identifiable vorticity signatures. Beyond this point, the discrepancy between the two estimation strategies becomes negligible.

A comparison of the reconstructed vorticity fields in Fig.~\ref{fig:AoA20_whole_plane} shows that, across all three time instants, $t \in \{1.6, 2.0, 2.6 \}$, the mean vorticity estimated using combined surface pressure and vorticity measurements ($\vor + \pres$) closely matches the reference field, with a portion of the error associated with the decoder reconstruction. This agreement includes both the gust structure beneath the airfoil and the downstream wake, with the estimator successfully reconstructing the negative lobe of the gust as it convects through the shadow region. The difference between the two reconstructions---shown in the third column as (Mean ($\vor + \pres$) - Mean ($\vor$))---is relatively small and remains localized near the airfoil. The color scale of this difference field spans only $1/15$ of the full vorticity magnitude, indicating that surface pressure measurements primarily refine the near-body flow rather than altering the global vortical structure. Consistently, posterior uncertainty, although small overall due to the large number of vorticity observations, is concentrated near the airfoil and within the shadow region during gust passage, while remaining low elsewhere. The close similarity between the vorticity reconstructions obtained with and without surface pressure measurements indicates that the analysis remains largely unchanged when pressure sensors are removed. This behavior is expected because the estimated state is a low-dimensional latent vector constrained to the learned autoencoder manifold. The dense vorticity measurements in the non-shadow region provide an overdetermined set of constraints on this reduced state, yielding a tightly concentrated posterior even in the absence of pressure data. The vorticity within the shadowed region is then inferred indirectly through correlations encoded by the decoder and the learned latent dynamics: once the latent state is identified from the observable region, the decoder completes the unobserved region with the most probable manifold-consistent reconstruction.
Importantly, even when the gust core resides beneath the airfoil, its influence is not confined locally in the latent representation. The induced flow leaves clear signatures in the observable flow field, rendering the latent dynamics identifiable without direct measurements in the shadow region.

\section{Conclusion}\label{sec:conclusion}
This study introduced a unified framework for sequential online estimation of unsteady aerodynamic flow fields, loads, and unknown body kinematics from sparse measurements in strongly disturbed environments. This framework is motivated by the practical challenges of full-field flow sensing in both laboratory and operational settings. The current approach integrates an offline stage of learning nonlinear reduced-order modeling with the physical ensemble-based Bayesian sequential filtering into a single online estimation architecture capable of operating under partial observability and spatially incomplete data. This framework is demonstrated on a challenging class of problems involving arbitrary vortical gust disturbances---with varying strength, size, and direction---interacting with airfoils undergoing randomly parameterized ramp pitch-up and hold maneuvers, highlighting its ability to simultaneously infer flow, loads, and body motion kinematics in highly transient and nonlinear settings.

The framework was systematically evaluated on two challenging problems. 
First, the primary contribution of this work lies in developing an estimation framework for problems in which a vortical gust interacts with a body with an unknown motion. For this configuration, we demonstrated, to the authors’ knowledge, for the first time, simultaneous fast estimation of unsteady flow fields, aerodynamic loads, and time-varying body kinematics using sparse surface pressure measurements alone. To enable this, a kinematics-aware flow autoencoder has been developed to learn a shared latent representation that jointly encodes the flow field and body kinematics, enabling the subsequent filtering step to be performed online in a computationally efficient and tractable reduced space. The estimation results captured both circulatory and non-circulatory load components, reproduced transient flow features during gust–airfoil interactions, and yielded narrow uncertainty bounds across a wide range of randomly sampled gust and pitch parameters. Observability in the weakly observed regions was further improved by augmenting the measurement vector with strategically placed vorticity slice measurements.

Second, we addressed experimentally realistic scenarios involving spatially incomplete flow observations, as commonly encountered in planar optical measurements with line-of-sight constraints. In these shadow-region configurations, typically beneath the airfoil, the estimator successfully reconstructed flow features---including gust-induced vortical structures---within these unobserved regions. The results revealed that dense vorticity measurements outside the shadowed region tightly constrain the low-dimensional latent state, allowing the decoder to complete the missing flow region in a physically consistent manner. In this planar vorticity measurements, the surface pressure sensors were shown to be mainly responsible for accurately capturing early non-circulatory lift responses.

Further insight into the observability of loads and motion kinematics through surface pressure measurements during the estimation process was obtained by performing an eigendecomposition of the state-space and observation-space Gramians. The leading observation modes capture the dominant directions through which pressure measurements constrain the latent space and were found to be primarily associated with lift-related adjustments. However, additional analysis of the sensitivities of drag, pitch angle, and angular velocity to surface pressure perturbations showed that intermediate observation modes also contribute significantly to their correction during the assimilation update. Consequently, truncating the update step based solely on the dominant modes of the latent-state and observation Gramians may remove latent directions that are weakly observable from pressure but remain important for estimating other quantities of interest. This observation is consistent with recent developments in balanced truncation for nonlinear model reduction, where retaining only the most energetic or dominant modes may neglect directions sensitive to specific outputs \citep{otto2023model}.

Overall, this work establishes a scalable and physically grounded framework for aerodynamic state estimation for moving bodies in disturbed environments, while providing a rigorous foundation for studying observability using surface pressure measurements. The proposed approach is directly applicable to sensing-based control, flow-aware decision making, and experimental diagnostics, where only sparse and indirect measurements are available, assuming that sufficient training data are available for learning the latent-space manifold on which the estimated flow's dynamics lie. In the present study, we focused on two-dimensional, low-Reynolds-number flows with simple prescribed body kinematics to systematically assess the performance of the framework on canonical unsteady aerodynamic problems. 
Future work will extend the methodology to three-dimensional configurations and higher Reynolds numbers, where increased flow complexity and stronger scale interactions pose additional challenges for real-time estimation. Beyond prescribed-motion cases, the framework naturally lends itself to fluid–structure interaction problems in which body motion is not imposed but instead emerges from the coupling between aerodynamic forces and structural dynamics. Further extensions will explore integration with reinforcement-learning and control architectures, enabling closed-loop, flow-aware autonomy in realistic aerodynamic environments.

\appendix
\section{Network architctures}
The detailed architecture of the kinematics-aware flow autoencoder, and latent dynamics are provided in Tables.~\ref{tab:network_blocks} and \ref{tab:network_blocks_forecast}, respectively.
\renewcommand{\thetable}{\thesection.\arabic{table}}
\setcounter{table}{0}

\renewcommand{\theequation}{\thesection.\arabic{equation}}
\setcounter{equation}{0}

\renewcommand{\thefigure}{\thesection.\arabic{figure}}
\setcounter{figure}{0}

\begin{table}[htbp]
  \centering
  \caption{Network architecture of the kinematics-aware flow autoencoder. The activation function is \texttt{Tanh}.}
  \label{tab:network_blocks}
  \renewcommand{\arraystretch}{3.0}  
  \small
  \begin{tabularx}{\textwidth}{|Y|Y||Y|Y||Y|Y||Y|Y|}
    \hline
    \multicolumn{4}{|c||}{\textbf{Encoder}} & 
    \multicolumn{4}{c|}{\textbf{Decoder}} \\
    \hline
    \multicolumn{2}{|c||}{\textbf{Vorticity}} &
    \multicolumn{2}{c||}{\textbf{Kinematics}} &
    \multicolumn{2}{c||}{\textbf{Vorticity}} &
    \multicolumn{2}{c|}{\textbf{Loads and kinematics}} \\
    \hline
    \textbf{Layer} & \textbf{Size} & 
    \textbf{Layer} & \textbf{Size} & 
    \textbf{Layer} & \textbf{Size} &
    \textbf{Layer} & \textbf{Size} \\
    \hline
    Input & \shortstack{(1, 120, \\240)} & Input & (2) & Input & ($n=$10) & Input & ($n=$10) \\
     \hline
    \shortstack{Conv2D\\(3,3,32)} & \shortstack{(32, 120, \\240)} & Dense & (64) & Dense & (128) & Dense & (64) \\
     \hline
    \shortstack{Conv2D\\(3,3,32)} & \shortstack{(32, 120, \\240)} & Dense & (64)  & Dense & (256) & Dense & (64) \\
     \hline
    \shortstack{MaxPool\\(2,2)} & \shortstack{(32, 60, \\120)} & & & Dense & (288) & Dense & (32) \\
     \hline
    \shortstack{Conv2D\\(3,3,16)} & \shortstack{(16, 60, \\120)}  & & & Reshape & \shortstack{(4, 6, \\12)} & \shortstack{Output 1\\($\pres$)} & (11) \\
     \hline
    \shortstack{Conv2D\\(3,3,16)} & \shortstack{(16, 60, \\120)}  & & & \shortstack{Conv2D\\(3,3,4)} & \shortstack{(4, 6, \\12)} & \shortstack{Output 2\\($\lift$)} & (1)  \\
     \hline
    \shortstack{MaxPool\\(2,2)} & \shortstack{(16, 30, \\60)}  & & & \shortstack{Conv2D\\(3,3,4)} & \shortstack{(4, 6, \\12)}  &  \shortstack{Output 3\\($\drag$)} & (1)  \\
     \hline
    \shortstack{Conv2D\\(3,3,8)} & \shortstack{(8, 30, \\60)}  & & & \shortstack{UpSample\\(5,5)} & \shortstack{(4, 30, \\60)} &  \shortstack{Output 4\\($\tilde{\ang}$)} & (1)  \\
     \hline
    \shortstack{Conv2D\\(3,3,8)} & \shortstack{(8, 30, \\60)}  & & & \shortstack{Conv2D\\(3,3,8)} & \shortstack{(8, 30, \\60)} &  \shortstack{Output 5\\($\tilde{\angv}$)} & (1)  \\
     \hline
    \shortstack{MaxPool\\(5,5)} & \shortstack{(8, 6, \\12)}  & & & \shortstack{Conv2D\\(3,3,8)} & \shortstack{(8, 30, \\60)} &  &  \\
     \hline
    \shortstack{Conv2D\\(3,3,4)} & \shortstack{(4, 6, \\12)}  & & & \shortstack{UpSample\\(2,2)} & \shortstack{(8, 60, \\120)} &  &  \\
     \hline
    \shortstack{Conv2D\\(3,3,4)} & \shortstack{(4, 6, \\12)}  & & & \shortstack{Conv2D\\(3,3,16)} & \shortstack{(16, 60, \\120)} &  &  \\
     \hline
    Reshape & (288) & & &  \shortstack{Conv2D\\(3,3,16)} & \shortstack{(16, 60, \\120)} & &  \\
     \hline
    Concat & \multicolumn{3}{c||}{288+64} & \shortstack{UpSample\\(2,2)} & \shortstack{(16, 120, \\240)} &  &  \\
     \hline
    Dense & \multicolumn{3}{c||}{256} & \shortstack{Conv2D\\(3,3,32)} & \shortstack{(32, 120, \\240)} &  &  \\
    \hline
    Dense & \multicolumn{3}{c||}{128} & \shortstack{Conv2D\\(3,3,32)} & \shortstack{(32, 120, \\240)} &  &  \\
    \hline
    Dense & \multicolumn{3}{c||}{$n=$10} & \shortstack{Conv2D\\(3,3,1)} & \shortstack{(1, 120, \\240)}  &  &  \\
    \hline
  \end{tabularx}
\end{table}

\begin{table}[htbp]
  \centering
  \caption{Network architecture of the Neural ODE. The activation function used is \texttt{Tanh}.}
  \label{tab:network_blocks_forecast}
  \renewcommand{\arraystretch}{2.0}  
  \small
  \begin{tabularx}{0.5\textwidth}{|Y|Y|}
    \hline
    \multicolumn{2}{|c|}{\textbf{Neural ODE}} \\
    \hline
    \textbf{Layer} & \textbf{Data Size} \\
    \hline
    Input & ($n=10$) \\
     \hline
    Dense & (128) \\
     \hline
    Dense & (256) \\
     \hline
    Dense & (128) \\
     \hline
    Output & ($n=10$) \\
     \hline
  \end{tabularx}
\end{table}

\subsection{Choice of activation function} \label{secap:activation}
Here, we discuss and justify the choice of the $\verb|tanh|$ activation function used in the present study. Although non-saturating activations---such as $\verb|Relu|$ or $\verb|Silu|$---can improve expressiveness, we found empirically that they may produce latent-to-physical mappings that are poorly conditioned for data assimilation. For these activations, the decoder Jacobian $J_{\decoder} (\lat) = \partial \decoder(\lat) / \partial \lat$ can exhibit large singular values along certain latent directions, meaning that small perturbations in $\lat$ can induce disproportionately large changes in reconstructed fields. Crucially, some of these high-gain directions may be weakly observable from surface pressure, allowing the EnKF to update $\lat$ in directions that reduce pressure misfit while generating unphysical flow reconstructions and unstable posterior ensembles.
By contrast, $\verb|Tanh|$ provides bounded, saturating nonlinearities,
\begin{equation}
    |\verb|Tanh|(\mathrm{x})| \leq 1, \ \ |\verb|Tanh|^\prime (\mathrm{x})| \leq 1,
\end{equation}
which implicitly moderates the sensitivity of the decoder mapping. In a deep network, the effective Lipschitz constant can be roughly bounded by
\begin{equation}\label{eq:lipschitz}
L \lesssim \prod_{\ell}|\weights_\ell|_2 \, \sup_{\mathrm{x}}|\verb|Tanh|'(\mathrm{x})|,
\end{equation}
which helps prevent the formation of extreme high-gain latent directions. As a result, the latent manifold becomes smoother and reconstructions vary more continuously under ensemble perturbations, leading to stable assimilation updates and physically plausible decoded fields across test cases. Based on these observations, we adopt $\verb|Tanh|$ activations throughout the autoencoder and the learned latent dynamical model.

\section{Sensitivity of the estimation to measurement noise variance} \label{secap:meas_noise}
In Section~\ref{sec:results}, the measurement noise variance for the surface pressure readings was set to $10^{-4}$. (Note that this variance corresponds to the non-dimensional pressure measurements, denoted by $\pres$.) It is nevertheless useful to examine the sensitivity of the reduced-order estimator to the assumed measurement noise variance. To this end, we evaluate the estimator under three noise levels, namely $10^{-4}$ (as used in the main text), $10^{-2}$, and $10^{-1}$, hereafter referred to as small, medium, and large noise levels, respectively. 

\begin{figure*}
\centering
\includegraphics[width=1.0\textwidth]{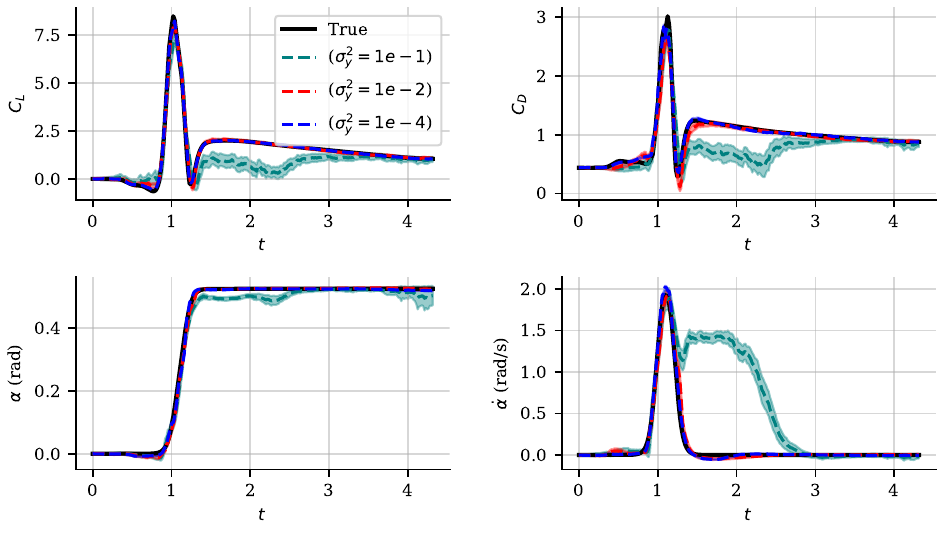}
\caption{\label{apfig:meas_noise_loads} Comparison of estimated lift under three different measurement noise variances. Results are shown for the case $\tramp=1.0$, $\kramp=1.125$, $D_y=0.58$, $\sigma=0.08c$, $y_o=0.07c$, and $t_o=0.5$.}
\end{figure*}

The results for the increased variance are shown in Fig.~\ref{apfig:meas_noise_loads}, where the predicted aerodynamic loads and body motion kinematics during the pitch-up maneuver are reported together with the corresponding uncertainty bounds. As expected, the uncertainty interval becomes noticeably wider for the case $\obsstd^2=10^{-1}$ reflecting the larger assumed measurement noise. In contrast, for $\obsstd^2=10^{-4}$ the uncertainty band is extremely narrow and barely visible in the figure.
Reduced measurement uncertainty, here either $\obsstd^2=10^{-2}$ or $10^{-4}$, leads to load and kinematics predictions that closely track the ground truth. However, the first deviation between the estimate and the true trajectory occurs for the case $\obsstd^2=10^{-1}$ sometime after the gust introduction here at $t_o=0.5$. At this stage, the filter must rely on the pressure measurements to detect the approaching gust and move the states toward their disturbed path. However, a large measurement noise variance weakens the state corrections during the update step (as analyzed later in this section), resulting in increased reliance on the forecast operator and continued evolution of the initial undisturbed flow until approximately $t \approx 0.8$, even after gust introduction. Consequently, because of the reduced correction in the case of the larger sensor noise, the filter requires several assimilation cycles---with relatively weak corrections---to recover and track the disturbed trajectory.
A second noticeable estimation error emerges in the high-noise case after $t=1.3$, as the gust convects downstream and the pressure-based observability diminishes, while the system trajectories relax toward their undisturbed equilibrium.

\begin{figure*}
\centering
\includegraphics[width=1.0\textwidth]{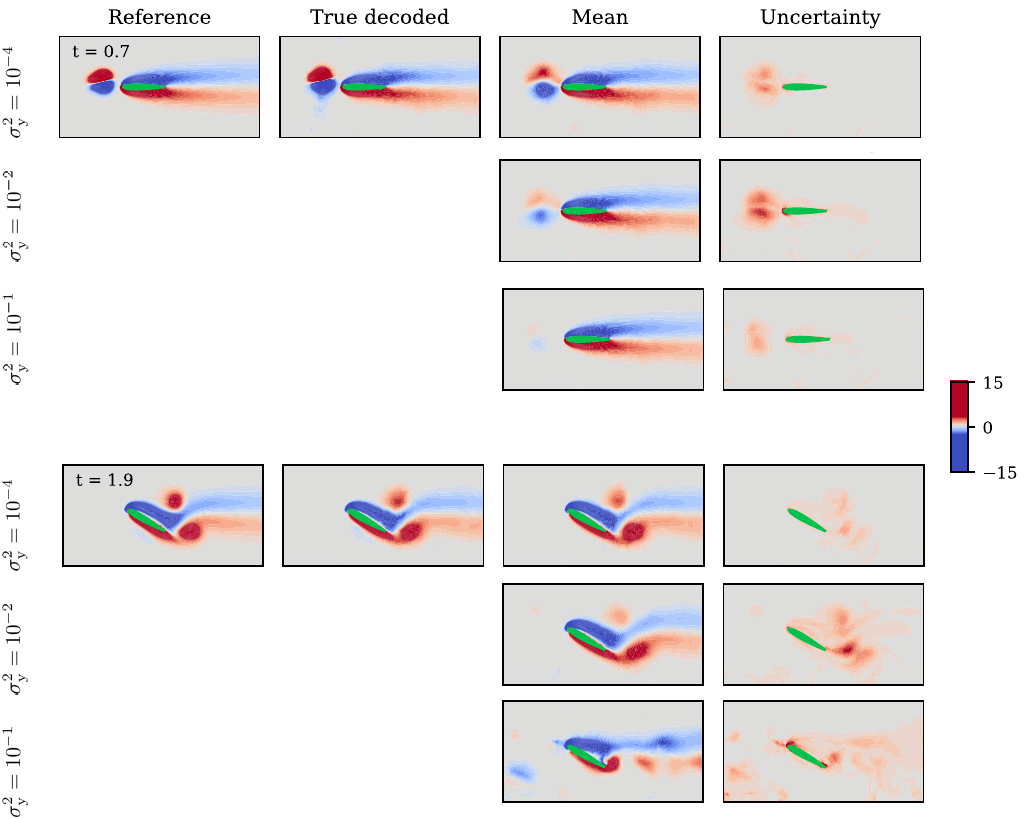}
\caption{\label{apfig:meas_noise_vor} Estimated vorticity field under three different measurement noise variances. Results are shown for the case $\tramp=1.0$, $\kramp=1.125$, $D_y=0.58$, $\sigma=0.08c$, $y_o=0.07c$, and $t_o=0.5$.}
\end{figure*}

Similarly, the effect of the measurement noise variance on the vorticity reconstruction is illustrated in Fig.~\ref{apfig:meas_noise_vor} at two representative times. The first instant, $t=0.7$, corresponds to an early stage following gust introduction, during which the disturbance remains upstream of the airfoil and the pitch-up maneuver has not yet commenced. The second instant, $t=1.9$, represents a later stage when the airfoil has already completed its pitch-up maneuver and marked by pronounced nonlinear interactions in the flow .
For low to moderate noise levels ($\obsstd^2=10^{-4}$ and $\obsstd^2=10^{-2}$), the reconstructed mean field closely matches the true decoded solution at both times, with the coherent vortical structures accurately captured and uncertainty remaining localized near dynamically active regions. This indicates that the estimator is largely insensitive to measurement noise within this regime.
As the noise variance increases to $\obsstd^2=10^{-1}$, noticeable degradation appears. At the earlier time ($t=0.7$), the estimator still captures the dominant flow topology, although ignoring the approaching gust at the upstream. At the later time ($t=1.9$), when the flow exhibits stronger nonlinear interactions and more distributed vortical structures, the reconstruction quality deteriorates significantly, with smeared vortices and elevated uncertainty across the wake.

These results indicate that the estimator remains robust and accurate for measurement noise covariances up to $\obsstd^2=10^{-2}$. The estimator is also observed to adapt its uncertainty bounds according to the magnitude of the measurement noise variance. As the noise level increases, the influence of measurement updates diminishes, and the estimator becomes more reliant on the learned forecast evolution, resulting in gradually increased deviation from the true states during long-time horizon. A more accurate forecast operator over long time horizons would likely mitigate the estimation errors observed at higher measurement noise levels.

\begin{figure*}
\centering
\includegraphics[width=1.0\textwidth]{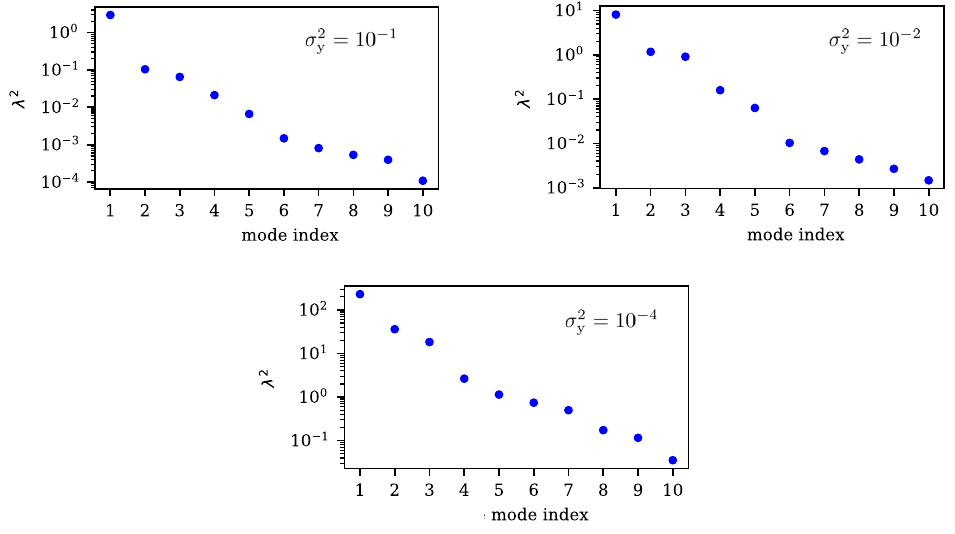}
\caption{\label{apfig:meas_noise_eigvals} Spectral plots of whitened Jacobian of observation operator at $t=0.7$ under three different measurement noise variances. Results are shown for the case $\tramp=1.0$, $\kramp=1.125$, $D_y=0.58$, $\sigma=0.08c$, $y_o=0.07c$, and $t_o=0.5$.}
\end{figure*}

The aforementioned stable behavior of the current estimator in increasing the sensor noise variance can be mathematically explained by examining how the Kalman update weights measurement information relative to the prior estimate in the whitened rotated space.
Assume a linearized form of the observation model defined in Eq.~\eqref{eq:observation}, expressed with respect to an arbitrary state vector $\state \in \mathbb{R}^n$---either the original high-dimensional state or its compressed latent representation $\lat$---as
\begin{equation}
    \y = \Obs \state + \obsnoise,
\end{equation}
where $\Obs$ denotes either a linear observation matrix or the Jacobian of a nonlinear observation operator (such as $\nabla \obs$ in the compressed space of the current study). In this section, we assume that $\state \sim \mathcal{N}(\pmb{0}, \sigmax)$ and that the observation error satisfies $\obsnoise \sim \mathcal{N}(\pmb{0}, \obsCov)$.
The whitened state and observation noise are defined as $\tilde{\state} = \sigmax^{-1/2} \state \in \mathbb{R}^n$ and $\tilde{\obsnoise} = \obsCov^{-1/2} \obsnoise \in \mathbb{R}^d$. Accordingly, the observation model in the whitened space becomes
\begin{equation}
    \tilde{\y} = \obsCov^{-1/2} \y = \tilde{\Obs} \tilde{\state} + \tilde{\obsnoise},
\end{equation}
where the whitened observation operator (or its tangent linear) is defined as $\tilde{\Obs} = \obsCov^{-1/2} \Obs \sigmax^{1/2} \in \mathbb{R}^{d \times n}$. Taking the singular value decomposition (SVD) of $\tilde{\Obs}$ gives
\begin{equation}
    \tilde{\Obs} = \pmb{U} \pmb{\Lambda} \pmb{V}^{\top},
\end{equation}
where, for the case $n<d$ considered here, $\pmb{U} \in \mathbb{R}^{d \times n}$ and $\pmb{V} \in \mathbb{R}^{n \times n}$ are the left and right singular vectors, respectively, and $\pmb{\Lambda} \in \mathbb{R}^{n \times n}$ is a diagonal matrix of singular values. As in the main text, we focus on the reduced-space setting with $d=11$ and $n=10$, so that $n<d$. 
We then project the state, observation noise, and observations onto these singular-vector bases as $\breve{\state} = \pmb{V}^{\top} \tilde{\state} \in \mathbb{R}^n$, $\breve{\obsnoise} = \pmb{U}^{\top} \tilde{\obsnoise} \in \mathbb{R}^n$, and $\breve{\y} = \pmb{U}^{\top} \tilde{\y} \in \mathbb{R}^n$. In the rotated space, the observation model becomes
\begin{equation} \label{apeq:rotated_obs}
    \breve{\y} = \pmb{\Lambda} \breve{\state} + \breve{\obsnoise}.
\end{equation}
Because both the prior covariance and the observation-noise covariance are identity matrices in the whitened space, and because $\pmb{\Lambda}$ is diagonal, inference in the rotated space is fully decoupled across singular directions.

Using the decoupled model in Eq.~\eqref{apeq:rotated_obs}, the Kalman gain formula of Eq.~\eqref{eq:kalman_gain_senkf} can be written in the rotated space as
\begin{equation}
    \breve{\gain} = \pmb{\Lambda} (\pmb{\Lambda}^2 + \pmb{I}_n)^{-1}.
\end{equation}
The corresponding transport map associated with Eq.~\eqref{eq:kalman_update} can then be written as
\begin{equation}
    \breve{T}(\breve{\state}; \breve{\y}^*_k) = \breve{\state} + \pmb{\Lambda}^2 (\pmb{\Lambda}^2 + \pmb{I}_n)^{-1} (\breve{\state}^* - \breve{\state}),
\end{equation}
where $\breve{\y}^* = \pmb{\Lambda} \breve{\state}^*$ and $\breve{\state}^*$ denotes the true state corresponding to the true observation in the rotated space. This expression corresponds to the noise-free case and is used here to clarify the mode-wise action of the transport map.

The factor $\pmb{\Lambda}^2 (\pmb{\Lambda}^2 + \pmb{I}_n)^{-1}$ is the \emph{state correction fraction}, which determines how much of the innovation is applied to the state in each singular direction. In each independent direction, this factor is
\begin{equation}
    \mathrm{scf} = \frac{\lambda_i^2}{\lambda_i^2 + 1} \qquad \ i = 1, 2, \cdots, n
\end{equation}
For large singular values, $\lambda_i \gg 1$, this factor approaches unity, indicating strongly observed directions in which the state is corrected almost fully toward the measurement-implied value. In contrast, when $\lambda_i \ll 1$, the state correction fraction approaches zero, indicating weakly observed directions along which the measurement update has little effect.

Under the three measurement noise variances considered in this section, $\obsstd^2=10^{-4}$, $10^{-2}$, and $10^{-1}$, the spectral values $\lambda_i^2$ are evaluated at $t=0.7$ when the disturbance is at upstream, and are shown in Fig.~\ref{apfig:meas_noise_eigvals}. The spectrum associated with the smallest measurement noise variance exhibits consistently larger eigenvalues, indicating stronger observability of the flow state from the available pressure sensors. Using the leading mode, $\lambda_1$, the corresponding state correction factor is found to be $\mathrm{scf}=0.997$ for $\obsstd^2=10^{-4}$, $\mathrm{scf}=0.90$ for $\obsstd^2=10^{-2}$, and $\mathrm{scf}=0.75$ for $\obsstd^2=10^{-1}$. Therefore, for $\obsstd^2=10^{-4}$ and $10^{-2}$, the filter update is governed almost entirely by the pressure measurements, and the prior state is corrected along the most observable directions. In contrast, for $\obsstd^2=10^{-1}$, the update reflects a more balanced compromise between the forecasted state and the measurement innovation. At this early stage, the forecast operator is unable to fully detect the approaching gust and continues to follow the undisturbed trajectory.

This behavior is especially important in the present aerodynamic problem, where disturbances may enter at arbitrary times relative to the motion maneuver, and the signature of the incoming gust is carried primarily by the surface pressure sensors. At such early stages, the forecast alone has limited ability to localize and reconstruct the approaching disturbance, whereas the measurement update provides the most direct information about its presence and strength. Accordingly, during these early regimes, stronger reliance on the pressure signatures leads to improved estimation and reconstruction, which explains the superior performance observed for the smallest measurement noise variance.

\section*{Acknowledgments}

The authors gratefully acknowledge the financial support provided by the National Science Foundation under award numbers 2247005 and 2247006.

\bibliographystyle{unsrtnat}
\bibliography{refs}

@article{mousavi2025sequential,
  title={{Sequential estimation of disturbed aerodynamic flows from sparse measurements via a reduced latent space}},
  author={Mousavi, Hanieh and Jones, Anya and Eldredge, Jeff},
  journal={arXiv preprint arXiv:2509.03795},
  year={2025}
}

@misc{eldredge2009computational,
	Author = {Eldredge, J. D. and Wang, C. and OL, M. V.},
	Howpublished = {39th AIAA Fluid Dynamics Conference. AIAA Paper 2009-3687},
	Title = {A Computational Study of a Canonical Pitch-Up, Pitch-Down Wing Maneuver},
	Year = {2009}}

@article{eldredge2022method,
  title={{A method of immersed layers on Cartesian grids, with application to incompressible flows}},
  author={Eldredge, Jeff D},
  journal={Journal of Computational Physics},
  volume={448},
  pages={110716},
  year={2022},
  publisher={Elsevier}
}

@article{fukami2023grasping,
  title={{Grasping extreme aerodynamics on a low-dimensional manifold}},
  author={Fukami, Kai and Taira, Kunihiko},
  journal={Nature Communications},
  volume={14},
  number={1},
  pages={6480},
  year={2023},
  publisher={Nature Publishing Group UK London}
}

@Article{eldredge2025practical,
  author  = {Eldredge, J. D. and Mousavi, H.},
  journal = {AIAA J.},
  title   = {Practical guide to flow estimation and uncertainty quantification of aerodynamic flows},
  year    = {2026},
  number  = {5},
  pages   = {2403--2423},
  volume  = {64},
  doi     = {10.2514/1.J066257}
}

@article{evensen2003ensemble,
  title={{The ensemble Kalman filter: Theoretical formulation and practical implementation}},
  author={Evensen, Geir},
  journal={Ocean dynamics},
  volume={53},
  number={4},
  pages={343--367},
  year={2003},
  publisher={Springer}
}

@book{evensen2009data,
  title={{Data assimilation: the ensemble Kalman filter}},
  author={Evensen, Geir},
  year={2009},
  publisher={Springer}
}

@article{le2021ensemble,
  title={{Ensemble Kalman filter for vortex models of disturbed aerodynamic flows}},
  author={Le Provost, Mathieu and Eldredge, Jeff D},
  journal={Physical Review Fluids},
  volume={6},
  number={5},
  pages={050506},
  year={2021},
  publisher={APS}
}

@techreport{stuart2015data,
  title={{Data assimilation: A mathematical introduction}},
  author={Stuart, Andrew and Zygalakis, Kostas},
  year={2015},
  institution={Oak Ridge National Lab.(ORNL), Oak Ridge, TN (United States)}
}

@article{nekkanti2023gappy,
  title={{Gappy spectral proper orthogonal decomposition}},
  author={Nekkanti, Akhil and Schmidt, Oliver T},
  journal={Journal of Computational Physics},
  volume={478},
  pages={111950},
  year={2023},
  publisher={Elsevier}
}

@article{alsalman2018training,
  title={{Training bioinspired sensors to classify flows}},
  author={Alsalman, Mohamad and Colvert, Brendan and Kanso, Eva},
  journal={Bioinspiration \& biomimetics},
  volume={14},
  number={1},
  pages={016009},
  year={2018},
  publisher={IOP Publishing}
}

@article{rezapour2026dynamic,
  title={Dynamic stall reattachment revisited},
  author={Rezapour, Sahar and Mulleners, Karen},
  journal={Journal of Fluid Mechanics},
  volume={1029},
  pages={A52},
  year={2026},
  publisher={Cambridge University Press}
}

@article{gementzopoulos2025flow,
  title={{Flow sensing through unsteady pressure measurements during transverse wing--gust encounters}},
  author={Gementzopoulos, Antonios and Wild, Oliver and Jones, Anya},
  journal={Experiments in Fluids},
  volume={66},
  number={3},
  pages={52},
  year={2025},
  publisher={Springer}
}

@article{medina2016leading,
  title={{Leading-edge vortex burst on a low-aspect-ratio rotating flat plate}},
  author={Medina, Albert and Jones, Anya R},
  journal={Physical Review Fluids},
  volume={1},
  number={4},
  pages={044501},
  year={2016},
  publisher={APS}
}

@article{le2022low,
  title={{A low-rank ensemble Kalman filter for elliptic observations}},
  author={Le Provost, Mathieu and Baptista, Ricardo and Marzouk, Youssef and Eldredge, Jeff D},
  journal={Proceedings of the Royal Society A},
  volume={478},
  number={2266},
  pages={20220182},
  year={2022},
  publisher={The Royal Society}
}

@article{tang2025neural,
  title={{Neural inference of fluid-structure interactions from sparse off-body measurements}},
  author={Tang, Rui and Zhou, Ke and Tan, Jifu and Grauer, Samuel J},
  journal={arXiv preprint arXiv:2506.23480},
  year={2025}
}

@article{zhu2025physics,
  title={{Physics-informed neural networks for hidden boundary detection and flow field reconstruction}},
  author={Zhu, Yongzheng and Chen, Weizheng and Deng, Jian and Bian, Xin},
  journal={arXiv preprint arXiv:2503.24074},
  year={2025}
}

@article{rodwell2024feel,
  title={{Feel the force: From local surface pressure measurement to flow reconstruction in fluid--structure interaction}},
  author={Rodwell, Colin and Sourav, Kumar and Tallapragada, Phanindra},
  journal={Physics of Fluids},
  volume={36},
  number={1},
  year={2024},
  publisher={AIP Publishing}
}

@article{liu2025attention,
  title={{Attention on flow control: transformer-based reinforcement learning for lift regulation in highly disturbed flows}},
  author={Liu, Zhecheng and Eldredge, Jeff D},
  journal={arXiv preprint arXiv:2506.10153},
  year={2025}
}

@article{chiereghin2019unsteady,
  title={{Unsteady lift and moment of a periodically plunging airfoil}},
  author={Chiereghin, Nicola and Cleaver, DJ and Gursul, Ismet},
  journal={AIAA Journal},
  volume={57},
  number={1},
  pages={208--222},
  year={2019},
  publisher={American Institute of Aeronautics and Astronautics}
}

@article{kurtulus2019unsteady,
  title={{Unsteady aerodynamics of a pitching NACA 0012 airfoil at low Reynolds number}},
  author={Kurtulus, Dilek Funda},
  journal={International Journal of Micro Air Vehicles},
  volume={11},
  pages={1756829319890609},
  year={2019},
  publisher={SAGE Publications Sage UK: London, England}
}

@article{mousavi2025low,
  title={{Low-order flow reconstruction and uncertainty quantification in disturbed aerodynamics using sparse pressure measurements}},
  author={Mousavi, Hanieh and Eldredge, Jeff D},
  journal={Journal of Fluid Mechanics},
  volume={1013},
  pages={A41},
  year={2025},
  publisher={Cambridge University Press}
}

@article{fukami2025extreme,
  title={{Extreme vortex-gust airfoil interactions at Reynolds number 5000}},
  author={Fukami, Kai and Smith, Luke and Taira, Kunihiko},
  journal={Physical Review Fluids},
  volume={10},
  number={8},
  pages={084703},
  year={2025},
  publisher={APS}
}

@article{de2024bio,
  title={{Bio-inspired flapping wing aerodynamics: a review}},
  author={De Manabendra, M and Sudhakar, Y and Gadde, Srinidhi and Shanmugam, Deepthi and Vengadesan, S},
  journal={Journal of the Indian Institute of Science},
  volume={104},
  number={1},
  pages={181--203},
  year={2024},
  publisher={Springer}
}

@article{renn2022machine,
  title={{Machine learning for flow-informed aerodynamic control in turbulent wind conditions}},
  author={Renn, Peter I and Gharib, Morteza},
  journal={Communications Engineering},
  volume={1},
  number={1},
  pages={45},
  year={2022},
  publisher={Nature Publishing Group UK London}
}

@article{bleckmann2009lateral,
  title={{Lateral line system of fish}},
  author={Bleckmann, Horst and Zelick, Randy},
  journal={Integrative zoology},
  volume={4},
  number={1},
  pages={13--25},
  year={2009},
  publisher={Wiley Online Library}
}

@article{mogdans2012coping,
  title={{Coping with flow: behavior, neurophysiology and modeling of the fish lateral line system}},
  author={Mogdans, Joachim and Bleckmann, Horst},
  journal={Biological cybernetics},
  volume={106},
  number={11},
  pages={627--642},
  year={2012},
  publisher={Springer}
}

@article{everson1995karhunen,
  title={{Karhunen--Loeve procedure for gappy data}},
  author={Everson, Richard and Sirovich, Lawrence},
  journal={Journal of the Optical Society of America A},
  volume={12},
  number={8},
  pages={1657--1664},
  year={1995},
  publisher={Optical Society of America}
}

@article{kaveh2026data,
  title={{Data assimilation in machine-learned reduced-order model of chaotic earthquake sequences}},
  author={Kaveh, Hojjat and Avouac, Jean Philippe and Stuart, Andrew M},
  journal={Geophysical Journal International},
  volume={244},
  number={2},
  pages={ggaf518},
  year={2026},
  publisher={Oxford University Press}
}

@article{cavanagh2024effect,
  title={Effect of sweep angle on three-dimensional vortex dynamics over plunging wings},
  author={Cavanagh, Alex and Bose, Chandan and Ramesh, Kiran},
  journal={Physics of Fluids},
  volume={36},
  number={11},
  year={2024},
  publisher={AIP Publishing}
}

@article{venturi2004gappy,
  title={{Gappy data and reconstruction procedures for flow past a cylinder}},
  author={Venturi, Daniele and Karniadakis, George Em},
  journal={Journal of Fluid Mechanics},
  volume={519},
  pages={315--336},
  year={2004},
  publisher={Cambridge University Press}
}

@article{luo2023reconstruction,
  title={Reconstruction of missing flow field from imperfect turbulent flows by machine learning},
  author={Luo, Zhaohui and Wang, Longyan and Xu, Jian and Wang, Zilu and Chen, Meng and Yuan, Jianping and Tan, Andy CC},
  journal={Physics of Fluids},
  volume={35},
  number={8},
  year={2023},
  publisher={AIP Publishing}
}

@article{aksoy2023reconstruction,
  title={Reconstruction of flow field with missing experimental data of a circular cylinder via machine learning algorithm},
  author={Aksoy, Muharrem Hilmi and Goktepeli, Ilker and Ispir, Murat and Cakan, Abdullah},
  journal={Physics of Fluids},
  volume={35},
  number={11},
  year={2023},
  publisher={Aip Publishing}
}

@article{luo2024deep,
  title={A deep learning framework for reconstructing experimental missing flow field of hydrofoil},
  author={Luo, Zhaohui and Wang, Longyan and Xu, Jian and Yuan, Jianping and Chen, Meng and Li, Yan and Tan, Andy CC},
  journal={Ocean Engineering},
  volume={293},
  pages={116605},
  year={2024},
  publisher={Elsevier}
}

@article{corkery2019quantification,
  title={{Quantification of added-mass effects using particle image velocimetry data for a translating and rotating flat plate}},
  author={Corkery, SJ and Babinsky, Holger and Graham, WR},
  journal={Journal of Fluid Mechanics},
  volume={870},
  pages={492--518},
  year={2019},
  publisher={Cambridge University Press}
}

@phdthesis{zaloglu2025,
  author       = {Zalo{\u{g}}lu, Berk},
  title        = {{Experimental Investigation of Gust Response in Flapping Wing Aerodynamics}},
  school       = {Istanbul Technical University},
  year         = {2025},
  type         = {{PhD thesis}}
}

@article{jones2022physics,
  title={Physics and modeling of large flow disturbances: discrete gust encounters for modern air vehicles},
  author={Jones, Anya R and Cetiner, Oksan and Smith, Marilyn J},
  journal={Annual Review of Fluid Mechanics},
  volume={54},
  number={1},
  pages={469--493},
  year={2022},
  publisher={Annual Reviews}
}

@article{biler2021experimental,
  title={Experimental investigation of transverse and vortex gust encounters at low Reynolds numbers},
  author={Biler, H{\"u}lya and Sedky, Girguis and Jones, Anya R and Saritas, Murat and Cetiner, Oksan},
  journal={AIAA Journal},
  volume={59},
  number={3},
  pages={786--799},
  year={2021},
  publisher={American Institute of Aeronautics and Astronautics}
}

@article{granlund2013unsteady,
  title={Unsteady pitching flat plates},
  author={Granlund, Kenneth O and Ol, Michael V and Bernal, Luis P},
  journal={Journal of Fluid Mechanics},
  volume={733},
  pages={R5},
  year={2013},
  publisher={Cambridge University Press}
}

@book{asch2016data,
  title={{Data Assimilation: Methods, Algorithms, and Applications}},
  author={Asch, Mark and Bocquet, Marc and Nodet, Ma{\"e}lle},
  year={2016},
  publisher={SIAM}
}

@article{loquercio2021learning,
  title={{Learning high-speed flight in the wild}},
  author={Loquercio, Antonio and Kaufmann, Elia and Ranftl, Ren{\'e} and M{\"u}ller, Matthias and Koltun, Vladlen and Scaramuzza, Davide},
  journal={Science Robotics},
  volume={6},
  number={59},
  pages={eabg5810},
  year={2021},
  publisher={American Association for the Advancement of Science}
}

@inproceedings{gao2018online,
  title={{Online safe trajectory generation for quadrotors using fast marching method and bernstein basis polynomial}},
  author={Gao, Fei and Wu, William and Lin, Yi and Shen, Shaojie},
  booktitle={2018 IEEE international conference on robotics and automation (ICRA)},
  pages={344--351},
  year={2018},
  organization={IEEE}
}

@article{lin2018autonomous,
  title={{Autonomous aerial navigation using monocular visual-inertial fusion}},
  author={Lin, Yi and Gao, Fei and Qin, Tong and Gao, Wenliang and Liu, Tianbo and Wu, William and Yang, Zhenfei and Shen, Shaojie},
  journal={Journal of Field Robotics},
  volume={35},
  number={1},
  pages={23--51},
  year={2018},
  publisher={Wiley Online Library}
}

@article{eldredge2019leading,
  title={{Leading-edge vortices: mechanics and modeling}},
  author={Eldredge, Jeff D and Jones, Anya R},
  journal={Annual Review of Fluid Mechanics},
  volume={51},
  number={1},
  pages={75--104},
  year={2019},
  publisher={Annual Reviews}
}

@article{bishop_adaptive_2001,
    title = {{Adaptive {Sampling} with the {Ensemble} {Transform} {Kalman} {Filter}. {Part} {I}: {Theoretical} {Aspects}}},
    volume = {129},
    issn = {0027-0644, 1520-0493},
    shorttitle = {Adaptive {Sampling} with the {Ensemble} {Transform} {Kalman} {Filter}. {Part} {I}},
    url = {http://journals.ametsoc.org/doi/10.1175/1520-0493(2001)129<0420:ASWTET>2.0.CO;2},
    doi = {10.1175/1520-0493(2001)129<0420:ASWTET>2.0.CO;2},
    language = {en},
    number = {3},
    urldate = {2025-11-11},
    journal = {Monthly Weather Review},
    author = {Bishop, Craig H. and Etherton, Brian J. and Majumdar, Sharanya J.},
    month = mar,
    year = {2001},
    pages = {420--436},
}

@article{otto2023model,
  title={{Model reduction for nonlinear systems by balanced truncation of state and gradient covariance}},
  author={Otto, Samuel E and Padovan, Alberto and Rowley, Clarence W},
  journal={SIAM Journal on Scientific Computing},
  volume={45},
  number={5},
  pages={A2325--A2355},
  year={2023},
  publisher={SIAM}
}

\end{document}